\documentclass{article}

\normalbaselines

\usepackage[T1]{fontenc}

\usepackage[all]{xy}

\PassOptionsToPackage{table}{xcolor}
\usepackage{graphics}
\usepackage{tikz}
\usepackage{url}
\usepackage{enumerate}

\usepackage{booktabs}
\usepackage{colortbl}
\usepackage{wrapfig}

\usepackage{times} 

\usepackage{amsmath}
\usepackage{amsthm}
\usepackage{amssymb}
\usepackage{xspace}
\usetikzlibrary{arrows,backgrounds,positioning}
\usetikzlibrary{snakes}

\DeclareMathOperator{\isPos}{pos}

\usepackage[ruled,vlined,linesnumbered,lined]{algorithm2e}

\def\calO{\mathcal{O}\xspace}
\newcommand{\game}{\mathcal{G}_\phi}
\newcommand{\NE}{Nash equilibrium\xspace}

\newcommand{\SE}{strong equilibrium\xspace}

\newcommand{\lasts}{\mathit{last}}

\usepackage{mathtools}
\usepackage{thmtools}
\usepackage{thm-restate}

\newcommand{\cdecor}[1]{{\bf #1}}
\def\ci{\cdecor{[i]}\xspace}
\def\cii{\cdecor{[ii]}\xspace}
\def\cio{\cdecor{[io]}\xspace}
\def\cioo{\cdecor{[ioo]}\xspace}
\def\cioi{\cdecor{[ioi]}\xspace}
\def\ciooi{\cdecor{[iooi]}\xspace}
\def\co{\cdecor{[o]}\xspace}
\def\coo{\cdecor{[oo]}\xspace}
\def\coi{\cdecor{[oi]}\xspace}
\def\cooi{\cdecor{[ooi]}\xspace}

\newtheorem{theorem}{Theorem}

\newtheorem{proposition}[theorem]{Proposition}
\newtheorem{claim}{Claim}
\newtheorem{lemma}[theorem]{Lemma}

\newtheorem{example}[theorem]{Example}
\newtheorem{corollary}[theorem]{Corollary}

\newtheorem{conjecture}{Conjecture}

\renewcommand {\qed} {\hfill \ensuremath{\Box} \\}

\newcommand{\payoff}{p}
\newcommand{\strprofile}{s}

\newcommand{\betredge}[1]{\betstep{#1}}
\newcommand{\betstep}[1]{\mbox{$\stackrel{#1}{\rightarrow}$}}

\newcommand{\guard}{\mathit{guard}}
\newcommand{\ochain}{open chain of cycles}

\def\ochains{open chains of cycles\xspace}

\newcommand{\nbr}{\mathit{NBR}}

\newcommand{\ssrem}[1]
{}

\newcommand{\lcnode}{v}

\newcommand{\ES}{\emptyset}

\newcounter{symbol}
\newcommand{\indexsyma}[1]%
{\stepcounter{symbol}\index{zzz1 \thesymbol @\protect#1}}
\newcommand{\indexsymb}[1]%
{\stepcounter{symbol}\index{zzz2 \thesymbol @\protect#1}}
\newcommand{\indexsymc}[1]%
{\stepcounter{symbol}\index{zzz3 \thesymbol @\protect#1}}
\newcommand{\indexsymd}[1]%
{\stepcounter{symbol}\index{zzz4 \thesymbol @\protect#1}}
\newcommand{\indexsyme}[1]%
{\stepcounter{symbol}\index{zzz5 \thesymbol @\protect#1}}

\newcommand{\bfe}[1]{\begin{bfseries}\emph{#1}\end{bfseries}\index{#1}}

\newcommand{\myra}{\mbox{$\:\rightarrow\:$}}

\newcommand{\lra}{\mbox{$\:\leftrightarrow\:$}}

\newcommand{\sse}{\mbox{$\:\subseteq\:$}}

\newcommand{\fa}{\mbox{$\forall$}}

\newcommand{\LL}{\mbox{$\ldots$}}

\newcommand{\C}[1]{\mbox{$\{{#1}\}$}}           

\newcommand{\NI}{\noindent}

\newcommand{\II}{\vspace{2 mm}}

\newcommand{\szkew}[1]{\relax \setbox0=\hbox{\kern -24pt $\displaystyle#1$\kern 0pt }%
\box0}
{\catcode`\@=11 \global\let\ifjusthvtest@=\iffalse}

\newcounter{oldmycaption}

\newif\iffull
\fulltrue

\renewcommand{\proof}[1]{\NI \mbox{#1}}
\renewcommand{\endproof}[1]{\mbox{#1}}

\begin{document}

\title{Coordination Games on Weighted Directed Graphs}

\author{Krzysztof R. Apt \\
       CWI, Amsterdam, The Netherlands, \\ MIMUW, University of Warsaw, Poland \\[2mm]
       Sunil Simon \\
       IIT Kanpur, Kanpur, India \\[2mm]
       Dominik Wojtczak \\
       University of Liverpool, Liverpool, U.K.
}


\date{}
\maketitle

\begin{abstract}
  We study strategic games on weighted directed graphs, in which the
  payoff of a player is defined as the sum of the weights on the edges
  from players who chose the same strategy, augmented by a fixed
  non-negative integer bonus for picking a given strategy.  These games
  capture the idea of coordination in the absence of globally common
  strategies.

  We identify natural classes of graphs for which finite
  improvement or coalition-improvement paths of polynomial length
  always exist, and, as a consequence, a (pure) \NE or a strong
  equilibrium can be found in polynomial time.  

  The considered classes of graphs are typical in network topologies:
  simple cycles correspond to the token ring local area networks,
  while open chains of simple cycles correspond to multiple
  independent rings topology from the recommendation G.8032v2 on the
  Ethernet ring protection switching.  For simple cycles these results
  are optimal in the sense that without the imposed conditions on the
  weights and bonuses a \NE may not even exist.

  Finally, we prove that the problem of determining the existence of a
  \NE or of a strong equilibrium in these games is NP-complete already
  for unweighted graphs and with no bonuses assumed.
  This implies that the same problems for polymatrix games are strongly NP-hard.
\end{abstract}







\section{Introduction}
\label{sec:intro}

\subsection{Background}

This paper is concerned with pure Nash equilibria in a natural
subclass of strategic form games.  Recall that a pure Nash equilibrium
of a strategic game is a joint strategy in which each player plays a
best response.  It is a natural solution concept which has been widely
used to reason about strategic interaction between rational
agents. Although Nash's theorem guarantees existence of a mixed
strategy Nash equilibrium for all finite games, pure Nash equilibria
need not always exist.  In various games, for instance Cournot
competition games or congestion games, pure Nash equilibria (from now,
just Nash equilibria) do exist and correspond to natural outcomes.

In many scenarios of strategic interaction, apart from the question of
the existence of Nash equilibria, an important concern is whether an
equilibrium can be efficiently computed.  In this context the concept
of an \emph{improvement path} is relevant.  These are maximal
paths constructed by starting at an arbitrary joint strategy and
allowing a single player who does not hold a best
response to switch to a better strategy at each stage. By definition, every finite
improvement path terminates in a Nash equilibrium.

In a seminal paper \cite{MS96}, Monderer and Shapley identified the class of finite games in
which every improvement path is guaranteed to be finite, and coined this property as the {\it finite improvement property} (FIP).
These are games with which one can associate a \emph{generalised ordinal
  potential}, a function on the set of joint strategies that properly
tracks the qualitative change in players' payoffs resulting from a
strategy change.  Thus the FIP not only guarantees the existence of
Nash equilibria but also ensures that it is possible to reach it from
any initial joint strategy by a simple update dynamics amounting to a
\emph{local search}.  This makes the FIP a desirable property. An
important class of games that have the FIP are the \emph{congestion
  games} that, as already noted in \cite{Ros73}, actually have an
\emph{exact potential}, a function that exactly tracks the
quantitative difference in players' payoffs.

However, the requirement that {\it every} improvement path is finite
is very strong and only a few classes of games have this property.
\cite{You93} proposed a weakening of the FIP that stipulates that from
any initial joint strategy only {\it some} improvement path is finite.
Games for which this property holds are called {\it weakly acyclic
  games}. So in weakly acyclic games Nash equilibria can be reached
through an appropriately chosen sequence of unilateral deviations of
players, irrespective of the starting joint strategy.

Although the existence of a finite improvement path guarantees the
existence of a Nash equilibrium, it does not necessarily result in an
efficient algorithm to compute it. In fact, in various
games, improvement paths can be exponentially long. \cite{FPT04} showed
that computing a Nash equilibrium in congestion games is PLS-complete.
Even in the class of symmetric network congestion games, for which it
is known that a Nash equilibrium can be efficiently computed
\cite{FPT04}, there are games in which some best response improvement
paths are exponentially long \cite{ARV06}. Thus identifying natural classes of
games in which starting from any joint strategy a Nash equilibrium can
be reached by an efficiently generated improvement path of polynomial
length is of obvious interest and is the focus of this paper.
	
\subsection{Motivation}

In game theory, coordination games are often used to model situations
of cooperation, where players can increase their payoffs by
coordinating on certain strategies. For two player games, this implies
that coordinating strategies constitute Nash equilibria. The main
characteristic of coordination is that players find it advantageous
that other players follow their choice. In this paper, we study a
simple class of multi-player coordination games, in which each player
can choose to coordinate his actions within a certain neighbourhood.
The neighbourhood structure is specified by a weighted directed graph,
the nodes of which are identified with the players.

Henceforth, we will refer to any strategy as a \bfe{colour}. The sets
of colours available to players are usually not mutually disjoint, as
otherwise players would not be able to coordinate on the same action.
Given a joint strategy, the payoff for a player is defined as the sum
of the weights of the incoming edges from other players who choose the
same colour plus a fixed bonus for picking this particular colour. We
refer to this subclass of strategic games as \bfe{weighted
  coordination games on graphs}, in short, just \bfe{coordination
  games}.
Coordination games capture the following key characteristics:

\begin{itemize}
\item {\it Join the crowd property:} the payoff of each player weakly
  increases when more players choose his strategy (this is because the weights
are assumed to be positive).

\item {\it Local dependency:} the payoff of each player depends only
  on the choices made by a certain group of players (namely the neighbours
  in the given weighted directed graph).

\item {\it Heterogeneous strategy sets:} players may have different strategy sets.

\item \emph{Individual preferences:} the (positive) bonuses express
  players' private preferences.

\end{itemize}

Coordination games constitutes a formal model to analyse strategic
interaction in situations where agents' benefit from aligning their
choices with other agents in their neighbourhood. Such circumstances
arise in various natural situations, for instance when clients have to
choose between multiple competing (for instance mobile phone)
providers offering similar services. It is often beneficial to choose
the same service provider as the one chosen by friends or
relatives. Thus, join the crowd property and local dependency
naturally hold. It is also natural to envisage that a provider imposes
some bounds on incentives that are provided. For instance, a
mobile phone operator might impose a cap on the number of free calls
and/or on the number of people with whom calls are free using its
network. Thus, weighted edges in the neighbourhood structure which
capture the quantitative ``influence'', in general, need not be
symmetric. Weighted directed edges are therefore appropriate to model
this general situation.

In this paper, we focus on the existence and efficient computation of
Nash equilibria in coordination games on specific directed graphs.
Given that players can try to coordinate their choice within a group,
it is also natural to consider a notion of equilibrium which takes
into account deviations by subsets of players. We therefore also study
the existence of strong equilibria, which are joint strategies from
which no subset of players can profitably deviate.  
We consider whether strong equilibria can be
efficiently computed by means of short improvement paths in which at
each stage all players in a group can profitably deviate. We call such
paths \emph{coalitional improvement paths}, in short
\emph{c-improvement paths}.

The coordination games studied here generalise the model 
introduced in \cite{ARSS14} and further studied in
\cite{AKRSS16}. In these works the neighbourhood structure is
represented by an unweighted and \emph{undirected} graph.  A switch to
\emph{directed} graphs turns out to be a major shift and leads to
fundamentally different results. For example, in the case of
undirected graphs, Nash equilibria always exist (in fact, these are exact potential games),
while even for simple directed graphs Nash
equilibria do not exist.  As a result both the structural results as
well as the techniques used here significantly differ from the ones in \cite{AKRSS16}.

A natural application of coordination games is in the analysis of
strategic behaviour in social networks. The threshold model
\cite{Gra78,AM11} in which members of the network are viewed as nodes
in a weighted graph, is one of the prevalent models used to reason
about social networks. Each node is associated with a threshold and a
node adopts an `item' (which can be a disease, trend, or a specific
product) when the total weight of incoming edges (or influence) from
the nodes that have already adopted this item exceeds its
threshold. The existence of directed edges is natural in such a
scenario, because the ``strength of influence'' captured by a quantitative value need not always be
symmetric between members in a social network.
When we omit bonuses, our coordination games become special
cases of the \emph{social network games} introduced and analysed in
\cite{SA15} provided one allows thresholds to be equal to 0.

\subsection{Related work}

The class of games that have the FIP, introduced in \cite{MS96}, was a
subject of extensive research. Prominent examples of such games are
congestion games.  Weakly acyclic games have received less attention,
but the interest in them is growing. \cite{Mil96} showed that although
congestion games with player specific payoff functions do not have the
FIP, they are weakly acyclic.  \cite{BV12} improved upon this result
by showing that a specific scheduling of players is sufficient to
construct a finite improvement path beginning at an arbitrary starting
point.  According to this scheduling the players are free to choose
their best response when updating their strategies.

Weak acyclicity of a game also ensures that certain modifications of
the traditional no-regret algorithm yield an almost sure convergence
to a Nash equilibrium \cite{MardenAS07}.  In \cite{ES11,ES14}, the authors show that 
specific Internet routing games are weakly
acyclic.  In turn, \cite{KL13} established that certain classes of
network creation games are weakly acyclic and moreover that a specific
scheduling of players can ensure that the resulting improvement path
converges to a Nash equilibrium in $\mathcal{O}(n \log n)$ steps.  Further, in \cite{MPRJ17}
the authors propose the use of weakly acyclic games
as a tool to analyse some iterative voting procedures.

Some structural results also exist.
\cite{FabrikantJS10} proved that the existence of a unique Nash
equilibrium in every subgame implies that the game is weakly
acyclic. A comprehensive classification of weakly acyclic games in
terms of schedulers is provided in \cite{AS12} and more extensively in
\cite{AS15}, where it was also shown that games solvable by means of
iterated elimination of never best responses to pure strategies are
weakly acyclic. Finally, \cite{Mil13} provided a characterization of weakly
acyclic games in terms of a weak potential and showed that every
finite extensive form game with perfect information is weakly acyclic.

As already mentioned, coordination games on unweighted and undirected graphs were
introduced and studied in \cite{AKRSS16}.  It was shown there that the
improvement paths are guaranteed to converge in polynomial number of
steps. Given this result, the study focused on the
analysis of strong equilibria and its variants. The authors also
provided bounds on the inefficiency of strong equilibria and
identified restrictions on the neighbourhood structure that ensure
efficient computation of strong equilibria.
These coordination games were augmented in \cite{RS15} by bonuses (which the authors call
\emph{individual preferences}). The authors studied the existence of
$\alpha$-approximate $k$-equilibria and their inefficiency
w.r.t.~social optima.  These equilibria are outcomes in which no group
of at most $k$ players can deviate in such a way that each member
increases his payoff by at least a factor $\alpha$.

The games we study here are related to various well-studied
classes of strategic form games. In particular, coordination games on graphs form a
natural subclass of {\it polymatrix games} \cite{Jan68}. These are
multi-player games where the players' utilities are pairwise
separable. Polymatrix games are well-studied and they include classes
of strategic form games with good computational properties like the
two-player zero-sum games.
\cite{simon2017constrained} studied the computational complexity of checking for the existence of constrained pure Nash equilibria in a subclass of polymatrix games defined on weighted directed graphs.
\cite{Hoefer2007} studied clustering games that are also polymatrix
games based on undirected graphs. In this setup each player has the
same set of strategies and as a result these games have, in contrast
to ours, the FIP.  A special class of polymatrix games was considered
in \cite{CD11}, which coincide with the coordination games on
undirected weighted graphs without bonuses.  The authors showed that
these games have an exact potential and that finding a pure Nash
equilibrium is PLS-complete. However, the proof of the latter result
crucially exploits the fact that the edge weights can be negative
(which captures anti-coordination behaviour). In
\cite{AS18} it was shown how coordination and anti-coordination on
simple cycles can be used to model and reason about the concept of
self-stabilization introduced in \cite{Dij74} one of the main
approaches to fault-tolerant computing.

When the graph is undirected and complete, coordination games on
graphs are special cases of the monotone increasing congestion games
that were studied in \cite{RT06}.

Another generalisation concerns distributed coalition formation
\cite{Haj06} where players have preferences over members of the same
coalition. Such a generalisation of polymatrix game over subsets of
players, called hypergraphical games, was introduced in
\cite{PR08}. Analysis of coalition formation games in the presence of
constraints on the number of coalitions that can be formed was
investigated in \cite{SHKW14}. \cite{SW17} studied a subclass of
hypergraphical games where the underlying group interactions are
restricted to coordination and anti-coordination. In this model,
players' utilities depend not just on the groups that are formed by
the strategic interaction, but also on the choice of action that the
members of the group decide to coordinate on.
It is shown that such games have a Nash equilibrium, which can be
computed in pseudo-polynomial time. Moreover, in the pure coordination
setting, when the game possesses a certain acyclic structure, strong
equilibria exist and can be computed in polynomial time. 

Coordination games on graphs are also related to \emph{additively
  separable hedonic games (ASHG)} \cite{Baner,Bogo}, which were
originally proposed in a cooperative game theory setting. In these
games players are the nodes of a weighted graph and can form
coalitions.  The payoff of a node is defined as the total weight of
all edges to neighbors that are in the same coalition.
The work on these games mostly focused on computational issues, see,
e.g., \cite{ABS10,ABS11,Aziz,Gair}.  

In \cite{AKRSS16} we also mentioned related work on strategic games
that involve colouring of the vertices of an undirected graph, in
relation to the vertex colouring problem.  In these games the players
are nodes in a graph that choose colours. However, the payoff function
differs from the one we consider here: it is $0$ if a neighbour
chooses the same colour and the number of nodes that chose the
same colour otherwise.  The reason is that these games are motivated by
the question of finding the chromatic number of a
graph. Representative references are \cite{PS08}, where it is shown
that an efficient local search algorithm can be used to compute a good
vertex colouring and \cite{EGM12}, where this work is extended by
analysing socially optimal outcomes and strong equilibria.
Further, strong and $k$-equilibria in strategic games on graphs were
also studied in Gourv\`es and Monnot \cite{GM09,GM10}. These games are
related to, respectively, the \texttt{MAX-CUT} and
\texttt{MAX-$k$-CUT} problems. These classes of games do not
satisfy the join the crowd property, so these results are not
comparable with ours.

\subsection{Our contributions}

In this paper we identify various natural classes of weighted directed graphs
for which the resulting games, possibly with bonuses, are weakly
acyclic.  Moreover, we prove that in these games, starting from any
arbitrary joint strategy, improvement paths of polynomial length can
be effectively constructed. So not only do these games have Nash
equilibria, but they can also be efficiently computed by a simple form
of local search.  Since coordination games on graphs are polymatrix
games, our results identify natural classes of polymatrix games in
which Nash equilibria are guaranteed to exist and can be computed
efficiently.

We first analyse coordination games on simple cycles. Even in
this limited setting, improvement paths of infinite length may
exist. However, we show that finite improvement paths always exist
when at most two nodes have bonuses or at most two edges have weights.
We also show that without these restrictions Nash equilibria may not
exist, so these results are optimal.  We then extend this setting to
\emph{open chains} of simple cycles, i.e., simple cycles that form a
chain and show the existence of finite improvement paths.

Most of our constructions involve a common, though increasingly more
complex, proof technique. In each case we identify a scheduling of
players that is easy to compute and such that, when combined with an
appropriate scheme to update strategies, 
guarantees that starting from an
arbitrary initial joint strategy, in the resulting improvement path,  
a Nash equilibrium is
reached in a polynomial number of steps.

We also study strong equilibria. In the restricted case of a weighted
directed acyclic graphs (DAGs) we show that strong equilibria can be
found along every coalitional improvement path. We also show that when
only two colours are used, the coordination games do not necessarily
have the FIP, but both Nash and strong equilibria can always be
reached starting from an arbitrary initial joint strategy by,
respectively, an improvement or a c-improvement path.

To deal with simple cycles we show that any finite improvement path
can be extended by just one profitable coalitional deviation to reach
a strong equilibrium.  This allows us to strengthen the results on the
existence of Nash equilibria to the case of strong equilibria.  We
also prove the existence of strong equilibria when the graphs are 
open chains of cycles. Finally, we show that in some coordination
games strong equilibria exist but cannot be reached from some initial
joint strategies by any c-improvement path.

Building upon these results we study the complexity of finding and
determining the existence of Nash equilibria and strong equilibria. In
particular we show that strong equilibrium in a coordination game on a
simple cycle can be computed in linear time.  
However, determining the existence of a Nash equilibrium even for games on unweighted graphs and without
bonuses, turns out to be NP-complete.

Table \ref{fig:summary} summarises our main results concerning the complexity of
finding Nash and strong equilibria.
For the complexity results we assume that all edge weights are natural numbers.
We list here respectively: the length of the shortest improvement
paths from an arbitrary initial joint strategy, the complexity of
finding a Nash equilibrium (abbreviated to NE), the length of the
shortest c-improvement paths starting from an arbitrary initial joint
strategy, and the complexity of finding a strong equilibrium
(abbreviated to SE). Here $n$ is the number of
nodes, $|E|$ the number of edges, and $l$ the number of colours.
In the case of open chain of cycles, $m$ denotes
the number of simple cycles in the chain and $v$ the number of nodes
in a simple cycle.

Most, though not all, results of this paper were reported earlier in
shortened versions, as two conference papers, \cite{ASW16} and
\cite{SW16}. Some of these results, notably on bounds on the length of
(c-)improvement paths, were improved.
\begin{table}[htbp]
  \centering
{\footnotesize
\newcolumntype{x}[1]{>{\centering\let\newline\\\arraybackslash\hspace{0pt}}p{#1}}
\setlength{\tabcolsep}{0.1cm}
\renewcommand{\arraystretch}{1.2}
\rowcolors{2}{white}{gray!20}
\newlength{\myl}
\settowidth{\myl}{weighted simple cycles+2 b}
\begin{tabular}{x{\myl}llll}
  \noalign{\global\belowrulesep=0.0ex}
  \hline
\noalign{\global\aboverulesep=0.0ex}
graph/bonus/colouring & improvement path & NE & c-impr. path & SE \\
\hline
weighted simple cycles with $\leq 1$ node with bonuses & $2n-1$ [Thm. \ref{thm:TARK15a}] & $\mathcal{O}(nl)$ [Thm. \ref{thm:cycles-complexity}]  & $2n$ [Cor. \ref{cor:TARK15}(i)] & $\mathcal{O}(nl)$ [Thm. \ref{thm:cycles-complexity}]   \\
simple cycles with bonuses  with $ \leq 1$ non-trivial weight & $3n -1$ [Thm.\ref{thm:TARK15}] & $\mathcal{O}(nl)$ [Thm. \ref{thm:cycles-complexity}]   & $3n$ [Cor. \ref{cor:TARK15}(ii)] & $\mathcal{O}(nl)$ [Thm. \ref{thm:cycles-complexity}]  \\
weighted simple cycles with $>2$ nodes with bonuses & \multicolumn{4}{c}{\NE may not exist [Example \ref{exa:simple_cycle}]}  \\
weighted simple cycles with $2$ nodes with bonuses & $3n$ [Thm. \ref{thm:cycle-2bonuses}] & $\mathcal{O}(nl)$ [Thm. \ref{thm:cycles-complexity}]  & $3n$ [Cor. \ref{cor:TARK15}(iii)] & $\mathcal{O}(nl)$ [Thm. \ref{thm:cycles-complexity}]  \\
simple cycles with bonuses and $2$ non-trivial weights & $4n-1$  [Thm. \ref{thm:cycle-2weights}] & $\mathcal{O}(nl)$ [Thm. \ref{thm:cycles-complexity}]  & $4n$ [Cor. \ref{cor:TARK15}(iiii)] & $\mathcal{O}(nl)$ [Thm. \ref{thm:cycles-complexity}]  \\
open chains of cycles  & $3vm^3$ [Thm. \ref{thm:necklace-noweight-nobonus}] & $\mathcal{O}(vm^3l)$ [Thm. \ref{thm:openchainNE-complexity}]  & $4vm^4$ [Thm. \ref{thm:se-open-chain}] & $\mathcal{O}(v^2m^5l)$ [Thm. \ref{thm:openchainSE-complexity}] \\
weighted DAGs with bonuses & $n-1$ [Thm. \ref{thm:DAG}] & $\mathcal{O}(nl + |E|)$ [Thm. \ref{thm:DAGs-complexity}] & $n-1$ [Thm. \ref{thm:DAG}] &  $\mathcal{O}(nl + |E|)$ [Thm. \ref{thm:DAGs-complexity}]  \\
two colours                     & $2n$ [Thm. \ref{thm:two-c1}] & $\mathcal{O}(n+|E|)$ [Thm. \ref{thm:two-se}]
 & $2n$ [Thm. \ref{thm:two-c}] &  $\mathcal{O}(n^2+ n|E|)$ [Thm. \ref{thm:two-se}] \\
\noalign{\global\aboverulesep=0.0ex}
\hline
\end{tabular}

\vskip0.5em
\caption{
\label{fig:summary}
Bounds on the length of the shortest improvement and c-improvement paths
for a given class of graphs or colouring and on the complexity of finding NE and SE. 
All edges are unweighted and there are no bonuses unless stated otherwise.
}
}
\end{table}

\subsection{Potential applications}

Coordination games constitute a natural and well-studied model that
represents various practical situations.  The class of games we study
in this paper models an extension of the coordination concept to a
network setting, where the network is represented as a weighted
directed graph, and where common strategies are not guaranteed to
exist while the payoffs functions take care of individual preferences.

The classes of graphs that we consider are frequently used as network
topologies. For example, the token ring local area networks are
organised in directed simple cycles, while the open chains of simple cycles
are supported by the recommendation G.8032v2 on the Ethernet ring
protection switching.\footnote{see
  \url{http://www.beldensolutions.com/en/Company/Press/PR103EN0609/index.phtml}}

The basic technique that we use to show finite convergence to Nash
equilibria is based on finite improvement paths of polynomial
length. The concept of an improvement path is fundamental in the study
of games but it also can be used to explain and analyse various real
world applications. One such example is the Border Gateway Protocol
(BGP) the purpose of which is to assign routes to the nodes of the
Internet and to use them for routing packets.
Over the years, there has been extensive research in the network
communications literature on how stable routing states are achieved
and maintained in BGP in spite of strategic concerns.  \cite{FP08} and
independently \cite{LSZ08} observed that the operation of the BGP can
be viewed as a best response dynamics in a natural class of routing
games and finite improvement paths that terminate in Nash equilibria
essentially translate to stable routing states. Following this
observation, \cite{ES14} presented a game
theoretic analysis of routing on the Internet in presence of
`misbehaving players' or backup edges.

Finally, coordination games on graphs are also relevant to cluster
analysis. Its main objective is to organise a set
of naturally related objects into groups according to some similarity
measure. When adopting the game-theoretic perspective one can view
possible cluster names as strategies and a satisfactory clustering of
the considered graph as an equilibrium in the coordination game
associated with the considered graph.  Clustering from a game
theoretic perspective (using evolutionary games) was among others
applied to car and pedestrian detection in images, and face
recognition, see \cite{PB14}. This approach was shown to perform very
well against the state of the art.

\subsection{Structure of the paper}  

In the next section we recall the relevant game-theoretic concepts and
the notions of (c-)impro\-ve\-ment paths, Nash and strong equilibria
on which we focus. In Section~\ref{sec:colouring} we introduce the
class of games which forms the subject of this paper.  The technical
presentation starts in Section~\ref{sec:simple-cycle} in which we
analyse the games the underlying graphs of which are (possibly
weighted) simple cycles. In Section~\ref{sec:necklace} we study open
chains of simple cycles.

Then, in Section \ref{sec:strong} we consider the problem of the
existence of strong equilibria.  Next, in Section
\ref{sec:directed-complexity}, we study the complexity of finding and
of determining the existence of Nash equilibria and strong equilibria.
We conclude by summarising in Section \ref{sec:conclusions} the
results and stating a natural open problem.

\section{Preliminaries}
\label{sec:prelim}

Throughout the paper $n > 1$ denotes the number of players.
A \bfe{strategic game} $\mathcal{G}=(S_1, \ldots, S_n,$ $p_1, \ldots,
p_n)$ for $n$ players, consists of a non-empty set $S_i$ of
\bfe{strategies} and a \bfe{payoff function} $p_i : S_1 \times \cdots
\times S_n \myra \mathbb{R}$, for each player $i$.  We denote $S_1
\times \cdots \times S_n$ by $S$, call each element $s \in S$ a
\bfe{joint strategy} and abbreviate the sequence $(s_{j})_{j \neq i}$
to $s_{-i}$. Occasionally we write $(s_i, s_{-i})$ instead of $s$.  We
call a strategy $s_i$ of player $i$ a \bfe{best response} to a joint
strategy $s_{-i}$ of his opponents if for all $ s'_i \in S_i$,
$p_i(s_i, s_{-i}) \geq p_i(s'_i, s_{-i})$.
A joint strategy $s$ is
called a \bfe{Nash equilibrium} if each $s_i$ is a best response to
$s_{-i}$.

Fix a strategic game $\mathcal{G}$.  We say that $\mathcal{G}$
satisfies the \bfe{positive population monotonicity (in short PPM)},
see \cite{KBW97}, if for all joint strategies $s$ and players $i, j$,
$p_i(s) \leq p_i(s_i, s_{-j})$.  (Note that $(s_i, s_{-j})$ refers to
the joint strategy in which player $j$ chooses $s_i$.) So if player
$j$ switches to player $i$'s strategy and the remaining players do not
change their strategies, then $i$'s payoff weakly increases.

Next, by a \bfe{polymatrix game}, see \cite{Jan68}, we mean a game
$(S_1,\ldots,S_n,\payoff_1,\ldots,\payoff_n)$ in which for all pairs
of players $i$ and $j$ there exists a \emph{partial} payoff function
$a^{ij}$ such that for any joint strategy
$\strprofile=(\strprofile_1,\ldots,\strprofile_n)$, the payoff of
player $i$ is given by
$\payoff_i(\strprofile):=\sum_{j \neq i}
a^{ij}(\strprofile_i,\strprofile_j)$.  So polymatrix games are
strategic games in which the influence of a strategy selected by a
player on the payoff of another player is always the same,
regardless of what strategies other players select.

We call a non-empty subset $K := \{k_1, \ldots, k_m\}$ of the set of
players $N:= \{1, \ldots, n\}$ a \bfe{coalition}. Given a joint
strategy $s$ we abbreviate the sequence $(s_{k_1}, \ldots, s_{k_m})$
of strategies to $s_K$ and $S_{k_1} \times \cdots \times S_{k_m}$ to
$S_{K}$. We occasionally write $(s_K, s_{-K})$ instead of $s$.

Given two joint strategies $s'$ and $s$ and a coalition $K$, we say
that $s'$ is a \bfe{deviation of the players in $K$} from $s$ if
$K = \{i \in N \mid s_i \neq s_i'\}$.  We denote this by
$s \betredge{K} s'$ and drop $K$ if it is a singleton. If in addition
$p_i(s') > p_i(s)$ holds for all $i \in K$, we say that the deviation
$s'$ from $s$ is \bfe{profitable} and say that $s \betredge{K} s'$ is
a \bfe{c-improvement step}. Further, we say that a coalition $K$
\bfe{can profitably deviate from $s$} if there exists a profitable
deviation of the players in $K$ from $s$.  Next, we call a joint
strategy $s$ a \bfe{k-equilibrium}, where $k \in \{1, \dots, n\}$, if
no coalition of at most $k$ players can profitably deviate from $s$.
Using this definition, a \bfe{Nash equilibrium} is a 1-equilibrium and
a \bfe{strong equilibrium}, see \cite{Aumann59}, is an
$n$-equilibrium.

A \bfe{coalitional improvement path}, in short a \bfe{c-improve\-ment
  path}, is a possibly infinite sequence $\rho=(s^1, s^2, \dots)$ of
joint strategies such that for every $k \geq 1$ there is a coalition
$K$ such that $s^{k} \betredge{K} s^{k+1}$ is a profitable deviation
of the players in $K$, with the property that if it is finite then it
cannot be extended. So if $\rho$ is finite then there is no
profitable deviation from the last element of the sequence that we denote
by $\lasts(\rho)$.  Clearly, if a c-improvement path is finite, its
last element is a strong equilibrium.

We say that $\mathcal{G}$ has the \bfe{finite c-improvement property}
(\bfe{c-FIP}) if every c-improvement path is finite. Further, we say
that the function $P: S \rightarrow A$, where $A$ is a set, is a
\bfe{generalised ordinal c-potential}, also called \bfe{generalised
  strong potential}, for $\mathcal{G}$ (see \cite{Harks13,Holzman97})
if for some strict partial ordering $(P(S), \succ)$ the fact that $s'$
is a profitable deviation of the players in some coalition from $s$
implies that $P(s') \succ P(s)$.  If a finite game admits a
generalised ordinal c-potential then it has the c-FIP.  The converse
also holds, see, e.g., \cite{AKRSS16}.

We say that $\mathcal{G}$ is \bfe{c-weakly acyclic} if for every joint
strategy there exists a finite c-improvement path that starts at
it. Thus games that are c-weakly acyclic have a strong
equilibrium.  We call a c-improvement path an \bfe{improvement path}
if each deviating coalition consists of one player. The notion of a
game having the \bfe{FIP} or being \bfe{weakly acyclic} is then
defined by referring to the improvement paths instead of c-improvement
paths.

In this paper we are interested in determining existence `short'
improvement and c-improvement paths starting from \emph{any} initial
joint strategy.  This motivates the following concept that we shall
extensively use.  We say that a game \bfe{ensures improvement paths of
  length $X$} (where $X$ can also be expressed using the
$\mathcal{O}(\cdot)$ function) if for each joint strategy there exists
an improvement path that starts at it and is of length (at most) $X$.
We use an analogous notion for the c-improvement paths. 

To find such 'short' (c-)improvement paths starting from an arbitrary
initial joint strategy we need to select the players in the right
order. This motivates the following notion.  By a \bfe{schedule} we
mean a finite or infinite sequence, each element of which is a player.
Let $\epsilon$ denote the empty sequence and $\emph{seq}:i$ the finite
sequence $\emph{seq}$ extended by $i$.  Given an initial joint
strategy $s$ a schedule generates an (not necessarily unique) initial
fragment of an improvement path defined inductively as follows:
\[
\begin{array}{l}
\emph{path}(s, \epsilon) := s, \\
\emph{path}(s, seq:i) := 
  \begin{cases}
    \emph{path}(s, seq)         & \mbox{if $i$ holds a best response in the last element of} \\
                                & \mbox{$\emph{path}(s, seq)$}, \\
    \emph{path}(s, seq)  \to s' & \mbox{otherwise}, \\
  \end{cases}
\end{array}
\]
where $s'$ is the result of updating the strategy of player $i$ in the
last element of $\emph{path}(s, seq)$ to a best response.

Sometimes we additionally specify how players update their strategies
to best responses, but even then the generated improvement paths do
not need to be unique. The process of selecting a strategy is always
linear in the number of strategies.  To show that a game ensures short
improvement paths we provide in each case an appropriate schedule.
Note that an infinite schedule can generate a finite improvement path,
which is the case when the last element of $\emph{path}(s, seq)$ is a
Nash equilibrium.

In the proofs we always mention the bounds on the improvement paths
but actually these are bounds on the relevant prefixes of the defined
schedules, which are always longer or of the same length.

\section{Coordination games on directed graphs}
\label{sec:colouring}

We now define the class of games we are interested in.  Fix a finite
set $M$ of $l$ colours. A \bfe{weighted directed graph} $(G,w)$ is a
pair, where $G=(V,E)$ is a directed graph without self loops and parallel
edges over the set of vertices $V=\{1,\ldots,n\}$ and $w$ is a function that
associates with each edge $e \in E$ a positive weight $w_e$.  We say
that a weight is \bfe{non-trivial} if it is different than 1.

Further, we say that a node $j$ is an \emph{in-neighbour} (from now on a
\bfe{neighbour}) of the node $i$ if there is an edge $j \to i$ in $E$.
We denote by $N_i$ the set of all neighbours of node $i$ in the graph
$G$.  A \bfe{colour assignment} is a function $C: V \to {\cal P}(M)$ which
assigns to each node of $G$ a non-empty set of colours.

We also introduce the concept of a \bfe{bonus}, which is a function
$\beta$ that assigns to each node $i$ and colour $c \in M$ a
non-negative integer $\beta(i,c)$.  When stating our results, bonuses
are assumed to be not present (or equivalently are assumed to be all
equal to 0), unless explicitly stated otherwise.  We say
that a bonus is \bfe{non-trivial} if it is different from the constant function 0.

Given a weighted graph $(G,w)$, a colour assignment $C$ and a bonus
function $\beta$ a strategic game $\mathcal{G}(G,w,C,\beta)$ is
defined as follows: 
\begin{itemize}
\item the players are the nodes,

\item the set of strategies of player (node) $i$ is the set of colours
  $C(i)$; we occasionally refer to the strategies as \bfe{colours},

\item the payoff function for player $i$ is
$p_i(s) = \sum_{j \in N_i,\, s_i = s_j} w_{j \to i} + \beta(i,s_i)$.
\end{itemize}

So each node simultaneously chooses a colour and the payoff to the
node is the sum of the weights of the edges from its neighbours that
chose its colour augmented by the bonus to the node for choosing its
colour.  We call these games \bfe{coordination games on weighted directed
  graphs}, from now on just \bfe{coordination games}.

Note that because the weights are non-negative each coordination game
satisfies the PPM.  When the weights of all the edges are 1, we are
dealing with a coordination game whose underlying graph is
unweighted. In this case, we simply drop the function $w$ from the
description of the game and drop the qualification `unweighted' when
referring to the graph.

Similarly, when all the bonuses are 0, we obtain a coordination game
without bonuses.  Likewise, in the description of such a game we omit
the function $\beta$. 
In a coordination game without bonuses when the underlying graph is
unweighted, each payoff function is simply defined by
$p_i(s) := |\{j \in N_i \mid s_i = s_j\}|$. Here is an example of such a game.

\begin{example} \label{exa:1}
\label{exa:payoff}
\rm
Consider the directed graph and the colour assignment
depicted in Figure~\ref{fig:example-graph} below.
Take in the corresponding coordination game the joint strategy that consists of the underlined colours. Then the payoffs are as follows:
\begin{itemize}
\item 0 for the nodes 1, 7, 8 and 9,
\item 1 for the nodes 2, 4, 5, 6, 
\item 2 for the node 3.
\end{itemize}

Note that this joint strategy is not a Nash equilibrium.
In fact, this game has no Nash equilibrium. To see this observe that
we only need to consider the strategies selected by the nodes 1, 2 and 3,
since each of the nodes 4, 5 and 6 always plays a best response by
selecting the strategy of its only predecessor and each of the nodes
7, 8, and 9 has just one strategy.

We now list all joint strategies for the nodes 1, 2 and 3 and in each
of them underline a strategy that is not a best response to the choice
of the other players: $(\underline{a},a,b)$, $(a,a,\underline{c})$,
$(a,c,\underline{b})$, $(a,\underline{c},c)$, $(b,\underline{a},b)$,
$(\underline{b},a,c)$, $(b,c,\underline{b})$ and
$(\underline{b},c,c)$.
\qed
\end{example}
\begin{figure}[htbp]
\centering
\tikzstyle{agent}=[circle,draw=black!80,thick, minimum size=2em,scale=0.8]
\begin{tikzpicture}[auto,>=latex',shorten >=1pt,on grid,scale=1.4]

\newdimen\R
\R=1.3cm
\newcommand{\llab}[1]{{\small $\{#1\}$}}
\draw (90: \R) node[agent,label=right:{\llab{a,\underline{b}}}] (1) {1};
\draw (90-120: \R) node[agent,label=right:{\llab{a,\underline{c}}}] (2) {2};
\draw (90-240: \R) node[agent,label=left:{\llab{b,\underline{c}}}] (3) {3};
\draw (30: \R) node[agent,label=right:\llab{a,\underline{b}}] (4) {4};
\draw (30-120: \R) node[agent,label=right:{\llab{a,\underline{c}}}] (5) {5};
\draw (30-240: \R) node[agent,label=left:{\llab{b,\underline{c}}}] (6) {6};
\draw (90: 1.9*\R) node[agent,label=right:{\llab{\underline{a}}}] (7) {7};
\draw (90-120: 2*\R) node[agent,label=right:{\llab{\underline{c}}}] (8) {8};
\draw (90-240: 2*\R) node[agent,label=left:{\llab{\underline{b}}}] (9) {9};
\foreach \x/\y in {1/2,2/3,3/1,1/4,4/2,2/5,5/3,3/6,6/1,7/1,8/2,9/3} {
    \draw[->] (\x) to (\y);    
}
\end{tikzpicture}
\caption{A coordination game with a selected joint strategy. \label{fig:example-graph}}
\end{figure}
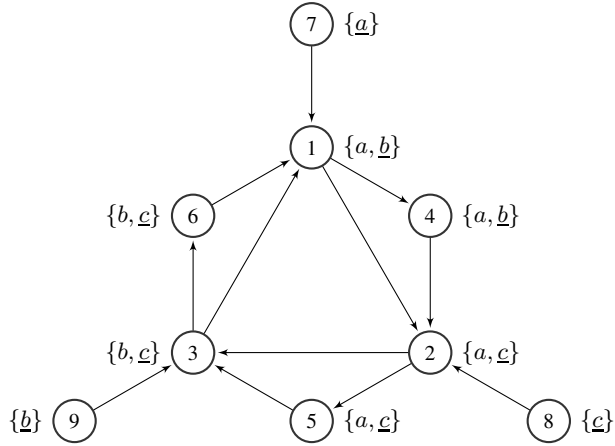

In the above game no bonuses are used and the edges in the underlying graph are unweighted. 
In Example \ref{ex:noNE} we exhibit a coordination game with bonuses which has a much 
simpler underlying graph with weighted edges and in which no Nash equilibrium exists.
The above example of course raises several questions, for instance, are there restricted classes of coordination games where a Nash equilibrium always exists, 
is the above example minimal in the
number of colours, does there exists coordination games that have a Nash
equilibrium but are not weakly acyclic, how difficult is it to determine
whether a Nash equilibrium exists, etc. We shall address these and
other questions in the rest of the paper.

\section{Simple cycles}
\label{sec:simple-cycle}

Given that coordination games need not always have a Nash equilibrium, 
we consider special graph structures to identify classes of games where a 
Nash equilibrium is guaranteed to exist. 
In this section we focus on simple
cycles. To fix the notation, suppose that the considered directed graph is
$1 \to 2 \to \ldots \to n \to 1$.  
We begin with the following simple example showing that the
coordination games on a simple cycle do not have the FIP.
Here and elsewhere, to increase readability, when presenting
profitable deviations we underline the strategies that were modified.

\begin{example} \label{exa:rotate}
\rm

Suppose $n \geq 3$.  Consider a coordination game on a simple cycle
where the nodes share at least two colours, say $a$ and $b$.  Take
the joint strategy $(a, b, \LL, b)$. Then both
$(a, \underline{b}, b, \LL, b) \to (a, a, b, \LL, b)$ and
$(\underline{a}, a, b, \LL, b) \to (b, a, b, \LL, b)$ are profitable
deviations.   After these two steps we obtain a joint strategy $(b, a, b, \LL, b)$
that is a rotation of the initial one. Iterating we obtain an infinite improvement
path.
\qed
\end{example}

On the other hand a weaker result holds. 

\begin{theorem}\label{thm:TARK15a}
Every coordination game on a weighted simple cycle in which at most
one node has bonuses ensures improvement paths of length $\leq 2n-1$.
\end{theorem}
\proof{\sc Proof.}
  First, assume that no node has bonuses.  Fix an initial joint
  strategy.  We construct the desired improvement path by scheduling
  the players in the round robin fashion, starting with player $1$.
  We prove that after at most two rounds we reach a Nash equilibrium.
  \II

\NI
{\em Phase 1.} This phase lasts at most $n-1$ steps. Each time we
select a player who does not hold a best response and update his
strategy to a best response. Such a modification affects only the
payoff of the successor player, so after we considered player $n-1$,
in the current joint strategy $s$ each of the players
$1, 2, \ldots, n-1$ holds a best response.

  If at this moment the current strategy of player $n$ is also a best
  response, then $s$ is a Nash equilibrium and the improvement path
  terminates. Otherwise we move to the next phase.  
\II

\NI
{\em Phase 2.} We repeat the same process as in Phase~1, but starting
with $s$ and player $n$.

By the definition of the game the property that at least $n-1$ players hold a
best response continues to hold for all consecutive joint strategies
and a Nash equilibrium is reached when the selected player holds a best response.

Suppose player $n$ switches to a strategy $c$.
Recall that $C(i)$ is the set of colours available to
player $i$.  Let 
\[
\begin{array}{l}
n_0 :=
  \begin{cases}
    n-1 \ \qquad \qquad \qquad \mbox{if $\fa i \in \{1, \LL, n-1\}: c \in C(i)$ and $s_i \neq c$} \\
    \mbox{$\min \{i \in \{1, \LL, n-1\} \mid c \not\in C(i) \mbox{ or } s_i = c\} - 1$ } \qquad \mbox{otherwise.} \\
  \end{cases}
\end{array}
\]

The improvement path terminates after the players $1, \LL, n_0$
successively switched to $c$ as at this moment player $n_0 + 1$ holds
a best response.

\II

Suppose that a node has bonuses. Then we rename the nodes so that this is
node $n$.  Then the argument used in reasoning about Phase 2 remains
correct.
\qed
\endproof
\II

As a side remark, note that the renaming of the players used at the
end of the above proof is necessary as otherwise the used schedule can
generate improvement paths that are longer than $2n-1$.

\begin{example}
\rm

Suppose that $n \geq 5$ and that the simple cycle is unweighted.
Assume that there are four colours $a, b, c, d$ and consider the
following colour assignment:
\[
C(1) = \LL = C(n-3) = C(n) = \C{a,b,c,d}, \ C(n-2) = \C{a, \overline{c}}, C(n-1) = \C{c,d},
\]
where the overline indicates the only positive bonus in the game.

Consider the joint strategy $(b, \LL, b, a, d, a)$. 
If we follow the clockwise schedule starting with player 1, there is
only one improvement path, namely
\[
\begin{array}{l}
  (\underline{b}, \LL, b, a, d, a) \to^* (a, \LL, a, a, d, \underline{a}) \to \\
(\underline{a}, \LL, a, a, d, d)   \to^* (d, \LL, d, \underline{a},d,d) \to (d, \LL, d, c, \underline{d}, d) \to (d,  \LL, d, c,c, \underline{d}) \to \\
(\underline{d}, \LL, d, c, c, c) \to^* (c, \LL, c, c, c, c).
\end{array}
\]

In each joint strategy we underlined the strategy of the scheduled
player from which he profitably deviates and each $\to^*$ refers to a
sequence of $n-3$ profitable deviations.  So this improvement path is
of length $3n - 5$ and thus longer than $2n-1$ since $n \geq 5$.
\qed
\end{example}

Further, the following result holds.

\begin{theorem}\label{thm:TARK15}
Every coordination game with bonuses on a simple cycle 
in which at most one edge has a non-trivial weight
ensures improvement paths of length $\leq 3n-1$.

\end{theorem}

\proof{\sc Proof.}
We first assume that no edge has a non-trivial weight.
  As in the proof of Theorem \ref{thm:TARK15a} we schedule the players
  clockwise starting with player $1$.  However, we are now more specific about the strategies to which the players switch.
Let $\mathit{MB}(i)$ be the set of available colours to player $i$
with the maximal bonus, i.e., 
\[
\mbox{$\mathit{MB}(i) := \{c \in C(i) \mid \beta(i,c) = \max_{d \in C(i)} \beta(i,d)\}.$}
\]  
  
Below we stipulate that whenever the selected player $i$ updates his
strategy
to a best response 
he always selects a strategy from $\mathit{MB}(i)$.  Note that this is
always possible, since the bonuses are non-negative integers.  Indeed,
suppose that the strategy of player's $i$ predecessor is $c$. If
$c \in \mathit{MB}(i)$, then player $i$ selects $c$ and otherwise he can
select an arbitrary strategy from $\mathit{MB}(i)$.
Fix an initial joint strategy.
\II

\NI 
{\em Phase 1.} This phase is the same as in the proof of Theorem
\ref{thm:TARK15a}, except the above proviso. So when this phase ends, 
the players $1, \ldots, n-1$ hold a best response.
If at this moment the current joint strategy $s$ is a Nash equilibrium,
the improvement path terminates. Otherwise we move to the next phase.
\II

\NI
{\em Phase 2.} We repeat the same process as in Phase~1, but starting
with $s$ and player $n$ and proceeding at most $n$ steps.  From now on
at each step at least $n-1$ players have a best response strategy.  So
if at a certain moment the scheduled player holds a best response, the
improvement path terminates.  Otherwise, the players $n, 1, \LL, n-1$
successively update their strategies and after $n$ steps we
move to the final phase.  
\II

\NI {\em Phase 3.} We repeat the same process as in Phase~2, again
starting with player $n$. In the previous phase each player updated
his strategy, so now in the initial joint strategy each player $i$
holds a strategy from $\mathit{MB}(i)$.  Hence each player can improve
his payoff only if he switches to the strategy selected by his
predecessor that also has the maximal bonus. Let $c$ be
the strategy to which player $n$ switches and let

\[
\begin{array}{l}
n_0 :=
  \begin{cases}
    n-1 \qquad \qquad \qquad \hfill \mbox{if $\fa i \in \{1, \LL, n-1\}: c \in MB(i)$ and $s_i \neq c$} \\
    \mbox{$\min \{i \in \{1, \LL, n-1\} \mid c \not\in MB(i) \mbox{ or } s_i = c\} - 1$ } \qquad
 \hfill \mbox{otherwise.} \\
  \end{cases}
\end{array}
\]

The improvement path terminates after the players $1, \LL, n_0$
successively switched to $c$ as at this moment player $n_0 + 1$ holds
a best response.

If some edge has a non-trivial weight then we rename the players so that
this edge is into the node $n$.  Notice that now we cannot require
that player $n$ selects a best response from $\mathit{MB}(n)$, since
the colour of his predecessor can yield a higher payoff due to the
presence of the weight. So we drop this requirement for node $n$ but
maintain it for the other nodes.

Then at the beginning of Phase 3 we can only claim that each player
$i \neq n$ holds a strategy from $\mathit{MB}(i)$, but this is
sufficient for the remainder of the proof.
\qed
\endproof
\II

We would like to generalise the above two results to coordination
games with bonuses on arbitrary weighted simple cycles.  However, the
following example shows that if we allow in a simple cycle non-trivial
weights on three edges and associate bonuses with three nodes then
some coordination games have no Nash equilibrium.

\begin{example} \label{exa:simple_cycle}
\label{ex:noNE}
\rm
Consider the weighted simple cycle and the colour assignment depicted
in Figure~\ref{fig:noNE}, where the overlined colours have bonus 1.

\begin{figure}[htbp]
\normalsize

\centering

\tikzstyle{agent}=[circle,draw=black!80,thick, minimum size=2em,scale=0.8]
\begin{tikzpicture}[auto,>=latex',shorten >=1pt,on grid]
\newdimen\R
\R=1.1cm
\newcommand{\llab}[1]{{\small $\{#1\}$}}
\draw (90: \R) node[agent,label=above:{\llab{\overline{a}, b}}] (1) {1};
\draw (90-120: \R) node[agent,label=right:{\llab{a,\overline{c}}}] (2) {2};
\draw (90-240: \R) node[agent,label=left:{\llab{\overline{b},c}}] (3) {3};
\foreach \x/\y in {1/2,2/3,3/1}{
    \draw[->] (\x) to node {{\small $2$}} (\y); 
}   
\end{tikzpicture}
\caption{A coordination game without a Nash equilibrium \label{fig:noNE}}
\end{figure}
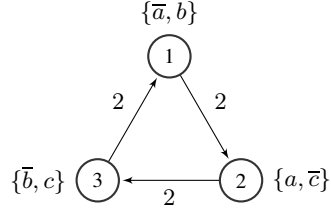

The resulting coordination game does not have a Nash equilibrium. The
list of joint strategies, each of them with an underlined strategy
that is not a best response to the choice of other players, is the
same as in Example \ref{exa:1}: $(\underline{a},a,b)$,
$(a,a,\underline{c})$, $(a,c,\underline{b})$, $(a,\underline{c},c)$,
$(b,\underline{a},b)$, $(\underline{b},a,c)$, $(b,c,\underline{b})$
and $(\underline{b},c,c)$.  In fact, the game considered in that example
simulates this game.  
\qed
\end{example}

In what follows we show that this counterexample is minimal in the
sense that if in a weighted simple cycle with bonuses at most two
nodes have bonuses or at most two edges have non-trivial weights, then
the coordination game has a Nash equilibrium. More precisely, we establish
the following two results.

\begin{theorem}
	\label{thm:cycle-2bonuses}
	Every coordination game on a weighted simple cycle in which 
	two nodes have bonuses ensures improvement paths of length $\leq 3n$.
	
\end{theorem}

\proof{\sc Proof.}
	Relabel the nodes if necessary so that one of the nodes which has 
	bonuses is node $1$. Let $k$ be the second node that has bonuses.
	Fix an initial joint strategy. We schedule, as before, 
        the players clockwise, starting with player $1$.

        \II
	\NI
        {\em Phase 1.} This phase lasts at most $n$ steps.  We
        repeatedly select the first player who does not hold a best
        response and update his strategy to a best response. A best
        response can be either the colour of the predecessor or, in
        the case of nodes $1$ and $k$ only, a colour with the maximal
        bonus.  In case of equal payoffs of these two options we give
        a preference to the former.
	As in the previous proofs, a strategy update of a given node,
        affects only the payoff of the successor node.  If at the end
        of Phase 1 the current strategy of player $1$ is also a best
        response, then we reached a Nash equilibrium and the
        improvement path terminates. Otherwise we move on to the next
        phase.
	
	\II
        \NI {\em Phase 2.} In this phase we perform at most
        two rounds of clockwise updates of all the nodes, starting at
        player $1$.
        We explicitly distinguish ten scenarios, which are defined as
        follows.  (They also play an important role in the proof of
        Theorem~\ref{thm:necklace-noweight-nobonus} in Section
        \ref{sec:necklace}.)  We focus on two types of strategy
        updates by the nodes with bonuses:

        \begin{itemize}
        \item  an update to an inner colour (recorded as
          \cdecor{i}), i.e., the colour of its predecessor, or

        \item an update to an outer colour (recorded as \cdecor{o}),
          i.e., one of the colours with a maximal bonus.

      \end{itemize}
      If a colour is both inner and outer, then we record it as
      \cdecor{i}.  An \emph{update scenario} is now a sequence of
      recordings of consecutive updates by the nodes with bonuses that
      is generated during the above two phases.

      One possible update scenario is \ciooi, which takes place when
      player $1$ first adopts the colour of its predecessor (\cdecor{i})
      and this colour then propagates until player $k$ is reached.  At
      this point player $k$ adopts a different colour with the maximal
      bonus (\cdecor{o}), and this colour propagates further until
      player $1$ is reached again.  Player $1$ then adopts a different
      colour with the maximal bonus (\cdecor{o}) which then propagates and
      is also adopted by player $k$ (\cdecor{i}). This propagation stops
      at a node $j$ lying between the nodes $k$ and $1$.
      At this point a Nash equilibrium is reached because
      player $j$ holds a best response and hence
      all players hold a best response.
	
      In general, an update scenario has to stop after an \cdecor{oi} or
      \cdecor{ii} is recorded, because then the same colour is
      propagated throughout the whole cycle and no new colour is
      introduced.  Moreover, an update string cannot contain
      \cdecor{ooo} as a subsequence, because then the third update to an
      outer colour would yield the same payoff as the first one, so
      it cannot be improving the payoff.  It is now easy to enumerate
      all update scenarios satisfying these two constraints and these
      are as follows: \co, \coi, \coo, \cooi, \ci, \cii, \cio, \cioi,
      \cioo, \ciooi. The only one of length 4 is the already
      considered update scenario \ciooi, which yields the longest
      sequence of profitable deviations in Phase 2, which is $2n$.
\qed
\endproof

Now consider coordination games on simple cycles with bonuses in which
two edges have non-trivial weights. The following example shows that
if we follow the clockwise schedule starting with player 1, then the
bound $3n$ given by Theorem~\ref{thm:cycle-2bonuses} does not need to
hold.

\begin{example}
  
\rm

Suppose that $n \geq 5$, the weights of the edges $n-3 \to n-2$ and
$n-1 \to n$ are 2 
and the weights of the other edges are 1. Let
$C = \C{a, b, c, d, e, f, g, h, i}$. Define the colour and the bonus
assignment as follows, where the overlined colours have bonus 1:
\[
\begin{array}{l}
C(1) = C \setminus \C{e}; \ \overline{f}, \overline{g}, \overline{i}, \\
C(2) = C \setminus \C{d}; \ \overline{e}, \overline{f}, \overline{g}, \overline{i}, \\
C(3) = \LL = C(n-3) = C, \\
C(n-2) = C \setminus \C{g,i}; \ \overline{h}, \\
C(n-1) = C \setminus \C{f}; \ \overline{g}, \overline{h}, \\
C(n) = C \setminus \C{h}; \ \overline{i},
\end{array}
\]

Consider now the joint strategy $(a, b, \LL, b,c,c,d)$.  If we follow
the clockwise schedule starting at player 1, we can generate the
following improvement path in which each player $i \neq n-2, n$ always
switches to a colour from $\mathit{MB}(i)$ (we cannot require it from players
$n-2$ and $n$ because the weights equal 2):
\[
\begin{array}{l}
(\underline{a},b, \LL, b,c,c,d) \to (d,\underline{b}, \LL, b,c,c,d) \to (d, \overline{e}, \underline{b}, \LL, b,c,c,d) \to (d, e, e, \underline{b}, \LL, b,c,c,d) \to^* \\
(\underline{d}, e, \LL, e,e,e) \to (\overline{f},\underline{e}, \LL, e,e,e) \to^* (f, \LL, f,\underline{e},e) \to (f, \LL, f,\overline{g},\underline{e}) \to \\
(\underline{f},\LL, f,g,g) \to^* (g,\LL, g, \underline{f},g,g) \to (g, \LL, g, \overline{h}, \underline{g}, g) \to (g, \LL, g, h, h, \underline{g}) \to \\
(\underline{g}, \LL, g, h,h, \overline{i}) \to^* (i, \LL, i, h, h,i).
\end{array}
\]
In each joint strategy we underlined the strategy of the scheduled
player from which he profitably deviates and overlined the first
occurrences of the newly introduced strategies.  Each $\to^*$ refers
to a sequence of $n-3$ profitable deviations.  So this improvement
path is of length $4n-3 > 3n-1$.  
\qed
\end{example}

However, a slightly larger bound can be established.

\begin{theorem}
\label{thm:cycle-2weights}
Every coordination game on a simple cycle with bonuses in which two
edges have non-trivial weights ensures improvement paths of length
$\leq 4n-1$.

\end{theorem}

\proof{\sc Proof.}
  Rename the nodes so that the edges with a non-trivial weight are
  into the nodes $k$ and $n$.  We stipulate that each player
  $i \neq k, n$ always selects a best response from the set
  $\mathit{MB}(i)$ of available colours to player $i$ with the maximal
  bonus. This is always possible for the reasons given in the proof of
  Theorem \ref{thm:TARK15}. As in the earlier proofs
  we construct the desired improvement path by scheduling
  the players clockwise, starting with player $1$. 
\II

\NI 
{\em Phase 1.} This phase lasts at most $2n-1$
steps. If this way we do not reach a Nash equilibrium we move to the next phase.
\II

\NI 
{\em Phase 2.} In this phase we continue the
clockwise strategy updates for all the nodes starting with player $n$.
We show that this can continue for at most two rounds.

In the second round of the previous phase each player $i \neq n$ updated
his strategy, so at the beginning of this phase each player $i \neq k, n$
holds a strategy from $\mathit{MB}(i)$.

We focus on the strategy updates by the nodes $k$ and
$n$.  To this end we reuse the reasoning used in the proof of Theorem
\ref{thm:cycle-2bonuses} that involves the analysis of the update scenarios.
So, as before, we distinguish between the updates of the nodes $k$ and
$n$ to an inner colour (recorded as \cdecor{i}) or to an outer colour
(recorded as \cdecor{o}) and consider the resulting update scenarios, so
sequences of \cdecor{i} and \cdecor{o}.

For the same reasons as before an update scenario has to stop after an
\cdecor{oi} or \cdecor{ii} is recorded, and it cannot contain contain
\cdecor{ooo} as a subsequence, as also here updates of a node to an
outer colour yield the same payoff.
Therefore the same argument shows that the longest
possible sequence of updates in this phase is $2n$.
\qed
\endproof

\section{Open chains of simple cycles}
\label{sec:necklace}

\newcommand{\nd}[2]{#1\!\!:\!\!#2}
\renewcommand{\nd}[2]{[#1,#2]}
\def\pl{\text{\sc +}\xspace}
\def\mn{\text{\sc --}\xspace}
\def\upl{\text{\sc u+}\xspace}
\def\umn{\text{\sc u--}\xspace}
\def\qm{\text{\sc ?}\xspace}
\def\calC{\mathcal{C}}

In this section we study directed graphs which consist of an open chain of
$m \geq 2$ simple cycles. For simplicity, we assume that all 
cycles have the same number of nodes denoted by $v$. The results we
show hold for arbitrary cycles as long as each cycle has at least 3
nodes.  Formally, for $j \in \{1,2, \ldots, m\}$, let $\mathcal{C}_j$
be the cycle
$\nd{j}{1} \to \nd{j}{2} \to \ldots \to \nd{j}{\lcnode} \to
\nd{j}{1}$. An \textit{\ochain{}} $\calC_1, \ldots, \calC_{m}$ is a
directed graph in which for all $j \in \{1,\ldots, m-1\}$ we have
$\nd{j}{1} = \nd{j+1}{k}$ for some $k \in \{2,\ldots,\lcnode\}$.  In
other words, it consists of a sequence of $m$ cycles such that any two
consecutive cycles have exactly one node in common.

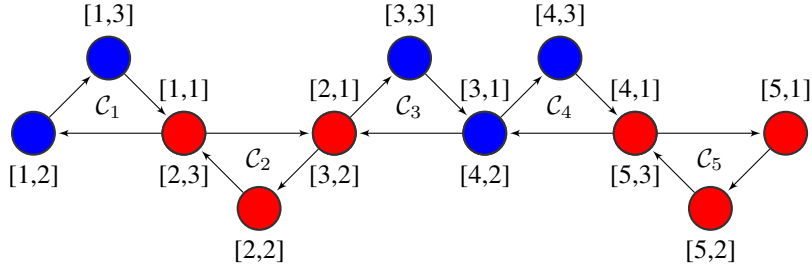
\begin{figure}[htbp]
    \centering
    \tikzstyle{agent}=[circle,draw=black!80,thick, minimum size=1.8em,scale=0.9]
    \begin{tikzpicture}[auto,>=latex',shorten >=1pt,on grid]
    \R=1.2cm
    \newcommand{\llab}[1]{{\small $\{#1\}$}}

    \foreach \x in {1,3,4}{
        \draw (2*\x+2,0) node[fill=red,agent,label={above:\nd{\x}{1}}] (3*\x+1) {};
        \draw (2*\x,0) node[fill=blue,agent,label={below:\nd{\x}{2}}] (3*\x+2) {};
        \draw (2*\x+1,1) node[fill=blue,agent,label={\nd{\x}{3}}] (3*\x+3) {};
        \draw (2*\x+1,0.35) node[] () {$\mathcal{C}_\x$};
    }
    \foreach \x in {2,5}{
        \draw (2*\x+2,0) node[fill=red,agent,label={above:\nd{\x}{1}}] (3*\x+1) {};
        \draw (2*\x,0) node[fill=red,agent,label={below:\nd{\x}{3}}] (3*\x+2) {};
        \draw (2*\x+1,-1) node[fill=red,agent,label={below:\nd{\x}{2}}] (3*\x+3) {};
        \draw (2*\x+1,-0.35) node[] () {$\mathcal{C}_\x$};
    }

    \foreach \x in {1,3,4}{
        \draw [->] (3*\x+1) to (3*\x+2);
        \draw [->] (3*\x+2) to (3*\x+3);
        \draw [->] (3*\x+3) to (3*\x+1);
    }
    
    \foreach \x in {2,5}{
        \draw [->] (3*\x+1) to (3*\x+3);
        \draw [->] (3*\x+3) to (3*\x+2);
        \draw [->] (3*\x+2) to (3*\x+1);
    }
    
    \end{tikzpicture}
    \caption{An open chain consisting of five cycles. Four nodes have double labels as they are link nodes. Each node can select either red or blue. The colouring of the nodes is an
      example of a joint strategy.}
    \label{fig:ochain}
\end{figure}

Any node that connects two cycles is called a {\em link node}.  The
node that connects $\mathcal{C}_j$ with $\mathcal{C}_{j+1}$, so
$\nd{j}{1}$, which is also $\nd{j+1}{k}$, is called an {\em up-link
  node} in $\mathcal{C}_j$ and, at the same time, a {\em down-link
  node} in $\mathcal{C}_{j+1}$.  The total number of nodes in such a
graph is $n = \lcnode m - (m-1)$.  Figure \ref{fig:ochain} depicts an
example of an open chain.

Throughout this section we assume a fixed coordination game on an
\ochain{} $\calC_1, \ldots, \calC_{m}$.  We prove that such a game
ensures improvement paths of polynomial length.  The main idea of our
construction is to build an improvement path by composing in an
appropriate way the improvement paths for the simple cycles that form the
open chain.

This is possible since, given a joint strategy, each cycle in the open
chain can be viewed as a single cycle with at most two bonuses for
which we know that an improvement path of length at most $3\lcnode$
exists due to Theorems~\ref{thm:TARK15a} and \ref{thm:cycle-2bonuses}.
This is because the only nodes that have indegree two are the link
nodes and given a joint strategy the edge to a link node $u$
from another cycle can be regarded as a bonus of $1$ for the colour of
the predecessor of $u$ in another cycle. More formally, for a given
joint strategy $s$ and a cycle $\calC_j$, we define the bonus function
$\beta_j^s(u,c)$
as follows:
\[
\begin{array}{l}
\beta_j^s(u,c) :=
  \begin{cases}
    1         & \mbox{if $u$ is a link node and $c = s(v)$,} \\[-1mm]
              & \mbox{where the node $v$ belongs to $\calC_{j-1}$ or to $\calC_{j+1}$ and $v \to u$ is an edge} \\
    0 & \mbox{otherwise.} \\
  \end{cases}
\end{array}
\]

Further, to each improvement path $\chi$ in the coordination game on
$\mathcal{C}_{j}$ with the bonus function $\beta^{s}_{j}$ there
corresponds a unique initial segment $\bar{\chi}$ of an improvement
path in the coordination game on the open chain
$\calC_1, \ldots, \calC_{m}$.
The following lemma will be useful a number of times.

\begin{lemma} \label{lem:payoff} Consider a coordination game on an
  open chain and a joint strategy $s$.  Each node with payoff $\geq 1$
  in $s$ plays a best response.  This also holds for coordination
  games on a simple cycle in which each node has at most one bonus
  equal to 1 and all other bonuses are 0.
\end{lemma}

\proof{\sc Proof.}
  The claim obviously holds for all nodes with the maximum possible payoff. Note that in the graphs considered here, for each node there are at most two colours that can give a payoff of $1$. These are the colours of the predecessors of a link node in an open chain, and the node's predecessor and the unique colour with bonus equal 1 in a simple cycle. The only possibility for such nodes to get payoff $2$ is if both of these colours coincide, which only depends on the colour(s) selected by its predecessor(s). Therefore, it is not possible for a node with a payoff of $1$ to unilaterally improve its payoff further.
\qed
\endproof

We claim that Algorithm \ref{alg:necklace} below finds an improvement
path of polynomial length.  It repeatedly tries to correct the cycle
with the least index in which some node does not play a best
response.

To express this procedure we use the constructions explained in the
proofs of Theorems~\ref{thm:TARK15a} and
\ref{thm:cycle-2bonuses}. Further, for a joint strategy $s$ that is
not a Nash equilibrium we denote by $\nbr(s)$ the least
$j \in \{1,\ldots,m\}$ such that some node in $\mathcal{C}_j$ does not
play a best response in $s$.  In the example given in Figure
\ref{fig:ochain} we have $\nbr(s) = 1$.

\begin{algorithm}
  \caption{\label{alg:necklace}}
  \SetNlSkip{1em}
	\KwIn{A coordination game on an open chain of cycles
          $\calC_1, \ldots, \calC_{m}$ and an initial joint strategy $s_0$.}  
	\KwOut{A finite improvement path starting at $s_0$.}
	$\rho:=s_0$;
        
	$s := \lasts(\rho)$;
        
	\While{$s$ is not a Nash equilibrium}{

		$j:=\nbr(s)$; 

                $\hat{s}:=$ the restriction of $s$ to the nodes of $\mathcal{C}_{j}$;
                
		$\chi:=$ the improvement path constructed in the proof of
                Theorem~\ref{thm:TARK15a} or \ref{thm:cycle-2bonuses}
                for the coordination game on $\mathcal{C}_{j}$ with
                the bonus function $\beta^{s}_{j}$, starting at
                $\hat{s}$;
		
		$\rho := \rho \bar{\chi}$;

                $s := \lasts(\rho)$}
	
	{\bf return} $\rho$.
\end{algorithm}

\def\grade{\textit{grade}}

The execution of this algorithm, when dealing with a cycle
$\mathcal{C}_j$, may `destabilise' some lower cycles, and hence may
require going back and forth along the sequence of cycles. In other words,
the value of $j$ may fluctuate.  However, we can identify
the minimum value below which $j$ cannot drop.

To see this we introduce the following notion.  Given a joint
strategy $s$ we assign to every cycle $\mathcal{C}_j$ one out of five
possible {\em grades}, \upl, \pl, \umn, \mn, and \qm, as follows:
\[
\begin{array}{l}
\grade^s(\mathcal{C}_j) :=
  \begin{cases}
        \upl         & \mbox{if all its nodes play their best response in $s$ and $s(\nd{j}{v}) = s(\nd{j}{1})$} \\
        \pl         & \mbox{if all its nodes play their best response in $s$ and $s(\nd{j}{v}) \neq s(\nd{j}{1})$} \\
    \umn         & \mbox{if $\nd{j}{2}$ is the only node that does not play a best response in $s$} \\[-1mm]
    & \mbox{and $s(\nd{j}{v}) = s(\nd{j}{1})$} \\
        \mn         & \mbox{if $\nd{j}{2}$ is the only node that does not play a best response in $s$} \\[-1mm]
        & \mbox{and $s(\nd{j}{v}) \neq s(\nd{j}{1})$} \\
        \qm & \mbox{otherwise.} \\
  \end{cases}
\end{array}
\]
Thus the grade \qm means that for some $k \neq 2$ the node
$\nd{j}{k}$ does not play a best response in $s$.

The following observation clarifies the relevance of the grade \upl and
is useful for the subsequent considerations.

\begin{lemma} \label{lem:upl} Suppose that after line 4 of Algorithm
  \ref{alg:necklace} the grade of a cycle $\mathcal{C}_i$ given $s$ is
  \upl and $j > i$. Then from that moment on $j > i$ remains true and
  the grade of $\mathcal{C}_i$ remains \upl.
\end{lemma}

\proof{\sc Proof.}
  During each \textbf{while} loop iteration $j$ can drop at most by 1,
  so the grade of $\mathcal{C}_i$ could be modified only if eventually
  after line 4 $j = i+1$ holds. The initial grade \upl of
  $\mathcal{C}_i$ implies that initially the colours of the nodes
  $\nd{i}{1}$ and $\nd{i}{v}$ are the same, and consequently the
  payoff for the node $\nd{i}{1}$ is $\geq 1$ and it remains so
  whenever its other predecessor, belonging to $\calC_{j}$, switches
  to another colour.

  But $\nd{i}{1}$ is also the down-link node $\nd{j}{k}$ of
  $\calC_{j}$.  Hence by Lemma \ref{lem:payoff} the improvement path
  constructed in line 6 does not modify the colour of $\nd{j}{k}$,
  i.e., of the node $\nd{i}{1}$.  So the grade of
  $\mathcal{C}_i$ remains \upl and hence if the \textbf{while} loop
  does not terminate right away, $j$ increases after line 4.
\qed
\endproof

Further, let $\grade(s)$ be the sequence of grades given $s$
assigned to each cycle, i.e.,
\[
  \grade(s) := (\grade^s(\mathcal{C}_1), \dots, \grade^s(\mathcal{C}_m)).
  \]
For instance, $\grade(s) = (\mn, \upl, \qm, \pl, \upl)$ for
the game and joint strategy $s$ presented in Figure \ref{fig:ochain}.

Suppose that Algorithm \ref{alg:necklace} selects $j$ in line 4.  It
then constructs in line 6 the improvement path that starts in
$\hat{s}$ defined in line 5, for the coordination game with bonuses on
the cycle $\calC_j$, as described in the proofs of
Theorems~\ref{thm:TARK15a} or \ref{thm:cycle-2bonuses}.  We now
explain how this can change $\grade(s)$.  Note that only the grades of
$\calC_j$ and its adjacent cycles $\calC_{j-1}$ and $\calC_{j+1}$ (if
they exist) can be affected.

\begin{restatable}{lemma}{lemgrades}
\label{lem:grades}
  The improvement path constructed in line 6 of Algorithm \ref{alg:necklace}
  modifies the grades of $\calC_j$ and its adjacent cycles $\calC_{j-1}$
  and $\calC_{j+1}$, if they exist, as explained in Figures \ref{tab:first},
  \ref{tab:umn}, \ref{tab:mn}, \ref{tab:qm}, and \ref{tab:last} below.
  \end{restatable}

\proof{\sc Proof.}
  We begin with some remarks and explanations.  $\nbr(s)$ returns the
  least index $j$ of a cycle with a node that does not play a best
  response.  So the initial grade of the cycle $\calC_{j-1}$, if
  it exists, is $\pl$ or $\upl$ and the initial grade of the cycle
  $\calC_j$ is \umn, \mn, or \qm.  Moreover, the grade of $\calC_j$
  can only change to \pl or \upl, because after line 6 all nodes in
  $\calC_j$ play a best response.  These observations allow us to
  limit the number of considered cases.

  In the presented tables we list above the horizontal bar the initial
  situation for the discussed cycles and under the bar one or
  more outcomes that can arise. Further, the initial grade of
  $\mathcal{C}_{j+1}$ is a parameter $x$.  If there are several
  options for the new grade of a given cycle, these are separated by
  /. Finally `any' is an abbreviation for
  \upl / \pl / \umn / \mn / \qm .

  Figure \ref{tab:first} corresponds to the case when $j=1$.  In turn,
  Figures \ref{tab:umn}, \ref{tab:mn}, and \ref{tab:qm} correspond to
  the cases when $1< j < m$ and initially the grade of $\calC_{j}$ is
  \umn, \mn, or \qm, respectively.  Finally, Figure \ref{tab:last}
  corresponds to the case when $j=m$.

  The cases considered in Figures \ref{tab:umn} and \ref{tab:mn} refer
  to the update scenarios defined in {\em Phase 2} in the proof of
  Theorem \ref{thm:cycle-2bonuses}. They are concerned with the relation of the
  colour of the up-link node in the cycle $\calC_j$ to the colour of
  its predecessor in this cycle.

  \begin{figure}[htbp]
	\centering
	\setlength{\cmidrulewidth}{1pt}
	\begin{tabular}{cc!{\qquad}cc!{\qquad}cc}
		\noalign{\global\belowrulesep=0.0ex}
		\toprule
		\noalign{\global\aboverulesep=0.0ex}
		\mn & $x$ & \umn & $x$ & \qm & $x$ \\ 
		\cmidrule(lr{2.4em}){1-2}
		\cmidrule(lr{2.4em}){3-4}
		\cmidrule(l){5-6}
		\pl/\upl & $x$ & \upl & $x$ & \pl/\upl & any \\ 
		\noalign{\global\aboverulesep=0.0ex}
		\bottomrule
	\end{tabular}
	\caption{Possible changes of the grades of $\mathcal{C}_j$ and $\mathcal{C}_{j+1}$
          when $j = 1$.}
	\label{tab:first}
\end{figure}
\begin{figure}[htbp]
	\centering
	\setlength{\cmidrulewidth}{1pt}
	\begin{tabular}{cccc!{\qquad}ccc}
		\noalign{\global\belowrulesep=0.0ex}
		\toprule
		\noalign{\global\aboverulesep=0.0ex}
		{\bf case} & \pl & \umn & $x$ & \upl & \umn & $x$ \\ 
		\cmidrule(lr{2.4em}){2-4}
		\cmidrule(l){5-7}
		\ci & \pl & \upl & $x$ & \upl & \upl & $x$ \\
		\cii & \pl/\mn/\upl/\umn & \upl & $x$ & \multicolumn{3}{c}{impossible} \\
		\cio & \umn/\upl & \pl/\upl & $x$ & \multicolumn{3}{c}{impossible} \\
		\cioi & \umn/\upl & \upl & any & \multicolumn{3}{c}{impossible} \\
		\cioo & \umn/\upl & \pl  & any & \multicolumn{3}{c}{impossible} \\
		\ciooi & \multicolumn{3}{c}{impossible} & \multicolumn{3}{c}{impossible} \\
		\noalign{\global\aboverulesep=0.0ex}
		\bottomrule
	\end{tabular}
	\caption{Possible changes of the grades of
          $\mathcal{C}_{j-1}$, $\mathcal{C}_j$, and $\mathcal{C}_{j+1}$
          when $1 < j < m$ and the grade of $\mathcal{C}_j$ is \umn.}
	\label{tab:umn}
\end{figure}
\begin{figure}[htbp]
	\centering
    \setlength{\cmidrulewidth}{1pt}
	\begin{tabular}{cccc!{\qquad}ccc}
		\noalign{\global\belowrulesep=0.0ex}
		\toprule
		\noalign{\global\aboverulesep=0.0ex}
		{\bf case} & \pl & \mn & $x$ & \upl & \mn & $x$ \\ 
		\cmidrule(lr{2.4em}){2-4}
		\cmidrule(l){5-7}
		\co & \pl & \pl & $x$ & \upl & \pl & $x$ \\
		\coi & \pl/\mn/\upl/\umn & \pl/\upl & $x$ & \multicolumn{3}{c}{impossible} \\
		\coo & \umn/\upl & \pl & $x$ & \multicolumn{3}{c}{impossible} \\
		\cooi & \umn/\upl & \upl & any & \multicolumn{3}{c}{impossible} \\
		\noalign{\global\aboverulesep=0.0ex}
		\bottomrule
	\end{tabular}
	\caption{Possible changes of the grades of
          $\mathcal{C}_{j-1}$, $\mathcal{C}_j$, and $\mathcal{C}_{j+1}$
          when $1 < j < m$ and the grade of $\mathcal{C}_j$ is \mn.}
	\label{tab:mn}
\end{figure}
\begin{figure}[htbp]
	\centering
	\setlength{\cmidrulewidth}{1pt}
	\begin{tabular}{ccc!{\qquad}ccc}
		\noalign{\global\belowrulesep=0.0ex}
		\toprule
		\noalign{\global\aboverulesep=0.0ex}
		\pl & \qm & $x$ & \upl & \qm & $x$ \\ 
		\cmidrule(lr{2.4em}){1-3}
		\cmidrule(l){4-6}
                \pl/\mn/\upl/\umn & \pl/\upl & any & \upl & \pl/\upl & any \\
		\noalign{\global\aboverulesep=0.0ex}
		\bottomrule
	\end{tabular}
	\caption{Possible changes of the grades of
          $\mathcal{C}_{j-1}$, $\mathcal{C}_j$, and $\mathcal{C}_{j+1}$
          when $1 < j < m$ and the grade of $\mathcal{C}_j$ is \qm.}
	\label{tab:qm}
\end{figure}
\begin{figure}[htbp]
	\centering
	\setlength{\cmidrulewidth}{1pt}
	\begin{tabular}{ccc!{\qquad}cc}
		\noalign{\global\belowrulesep=0.0ex}
		\toprule
		\noalign{\global\aboverulesep=0.0ex}
		{\bf case} & \pl & \umn & \upl & \umn  \\ 
		\cmidrule(lr{2.4em}){2-3}
		\cmidrule(l){4-5}
		\ci & \pl & \upl & \upl & \upl  \\
		\cii & \pl/\mn/\upl/\umn & \upl & \multicolumn{2}{c}{impossible} \\
		\cio & \umn/\upl & \pl/\upl & \multicolumn{2}{c}{impossible} \\
		\cioi & \umn/\upl & \upl  & \multicolumn{2}{c}{impossible} \\
		\noalign{\global\aboverulesep=0.0ex}
	\end{tabular}
        \quad
	\begin{tabular}{ccc!{\qquad}cc}
		\noalign{\global\belowrulesep=0.0ex}
		\toprule
		\noalign{\global\aboverulesep=0.0ex}
		{\bf case} & \pl & \mn &  \upl & \mn  \\ 
		\cmidrule(lr{2.4em}){2-3}
		\cmidrule(l){4-5}
                \co & \pl & \pl & \upl & \pl  \\
		\coi & \pl/\mn/\upl/\umn & \pl/\upl  & \multicolumn{2}{c}{impossible} \\
		\coo & \umn/\upl & \pl &  \multicolumn{2}{c}{impossible} \\
		\cooi & \umn/\upl & \upl & \multicolumn{2}{c}{impossible} \\
                \noalign{\global\aboverulesep=0.0ex}
		\bottomrule
	\end{tabular}
        
	\begin{tabular}{cc!{\qquad}cc}
		\noalign{\global\belowrulesep=0.0ex}
		\noalign{\global\aboverulesep=0.0ex}
		\pl & \qm & \upl & \qm  \\ 
		\cmidrule(lr{2.4em}){1-2}
		\cmidrule(l){3-4}
          \pl/\mn/\upl/\umn & \pl/\upl  & \upl & \pl/\upl  \\
		\noalign{\global\aboverulesep=0.0ex}
		\bottomrule
	\end{tabular}
	\caption{Possible changes of the grades of
          $\mathcal{C}_{j-1}$ and $\mathcal{C}_j$
          when $j=m$.}
	\label{tab:last}
\end{figure}

The justifications of these changes of the grades are lengthy and are provided in the appendix.
\qed
\endproof

Next, we introduce a progress measure $\mu$ defined
  on the current joint strategy that increases according to the
  lexicographic order each time the joint strategy $s$ is
  modified in line 8. In effect $\mu$ is a weak potential in the sense of
  \cite{Mil13}. $\mu(s)$ is a quadruple the definition of which uses the function $\nbr$ and
  two other functions that we now define.
  
Let $\guard(s)$ be the largest
$j \in \{1, \ldots, m\}$ such that given $s$ the grade of $\calC_j$ is
\upl and the grade of each cycle $\calC_1, \ldots, \calC_{j-1}$ is
either \pl or \upl. If no such $j$ exists, as it is the case in the
example given in Figure \ref{fig:ochain}, then we let $\guard(s)=0$.

\def\prefix{\textit{prefix}}

Further, let $\prefix(s)$ be the longest prefix of $\grade(s)$ such
that at most one of the grades it contains is \mn, \umn, or \qm.
Moreover, this prefix stops after a cycle with grade \qm.  For the
example given in Figure \ref{fig:ochain} we have
$\prefix(s) = (\mn, \upl)$.

Here is an example illustrating the introduced notions to which we shall return shortly.

\begin{example} \label{exa:openchain}
Suppose that $\grade(s_1) := (\pl, \upl, \upl, \pl, \mn, \umn, \qm)$ for a joint strategy $s_1$.
Then $\nbr(s_1) = 5$, $\guard(s_1) = 3$, and $\prefix(s_1) = (\pl, \upl,
\upl, \pl, \mn)$. Suppose that $\grade(s_2) := (\pl, \upl, \mn, \pl,
\upl, \umn, \umn, \qm)$ for a joint strategy $s_2$. Then $\nbr(s_2) = 3$, $\guard(s_2) = 2$, and
$\prefix(s_2) = (\pl, \upl, \mn, \pl, \upl)$.  \qed
\end{example}

\def\mytop{(m+1,0,0,0)}

We can now define $\mu(s)$.
First, we set $\mu(s) = \mytop$ if $s$ is a Nash equilibrium.
Otherwise
\[
\begin{array}{l}
\mu(s) :=
  \begin{cases}
    (\guard(s), 1, 0, -\nbr(s))        & \mbox{if $\prefix(s)$ contains \umn or if it contains 
      \upl} \\[-2mm]
                                       & \mbox{somewhere after \mn} \\
    (\guard(s), 0, |\prefix(s)|, -\nbr(s)) & \mbox{otherwise.} \\
  \end{cases}
\end{array}
\]

For example, for the joint strategies $s_1$ and $s_2$ used in Example \ref{exa:openchain}, we have $\mu(s_1) = (3,0,5,-5)$ and $\mu(s_2) = (2,1,0,-3)$ respectively.

To see the evolution of the progress measure $\mu(s)$ we
present in Figure \ref{fig:progress} an example run of Algorithm \ref{alg:necklace} on an open chain of eight cycles by recording at each step the corresponding changes of the grades and of the progress measure. It illustrates the fact that
during the execution of the algorithm the index of the first cycle with no Nash
equilibrium, i.e., the value of $\nbr(s)$, can arbitrarily decrease.

\begin{figure}[htbp]
    \centering
	\setlength{\cmidrulewidth}{1pt}
    \begin{tabular}{cccccccc!{\quad}c}
        \noalign{\global\belowrulesep=0.0ex}
        \toprule
        \noalign{\global\aboverulesep=0.0ex}
        \multicolumn{8}{c}{$\grade(s)$} & $\mu(s)$ \\
        \cmidrule{1-8}
        \pl & \pl & \pl & \pl & \qm & \upl & \upl & \mn & $(0, 0, 5, -5)$ \\
        \pl & \pl & \pl & \mn & \pl & \qm & \upl & \mn & $(0, 0, 5, -4)$ \\
        \pl & \pl & \mn & \pl & \pl & \qm & \upl & \mn & $(0, 0, 5, -3)$ \\
        \pl & \umn & \upl & \qm & \pl & \qm & \upl & \mn & $(0, 1, 0, -2)$ \\
        \umn & \pl & \qm & \qm & \pl & \qm & \upl & \mn & $(0, 1, 0, -1)$ \\
        \upl & \pl & \qm & \qm & \pl & \qm & \upl & \mn & $(1, 0, 3, -3)$ \\
        \upl & \pl & \upl & \qm & \pl & \qm & \upl & \mn & $(3, 0, 4, -4)$ \\
        \upl & \pl & \upl & \pl & \pl & \qm & \upl & \mn & $(3, 0, 6, -6)$ \\
        \upl & \pl & \upl & \pl & \pl & \upl & \upl & \mn & $(7, 0, 8, -8)$ \\
        \upl & \pl & \upl & \pl & \pl & \upl & \upl & \pl & $(9, 0, 0, 0)\phantom{-}$ \\
        \noalign{\global\aboverulesep=0.0ex}
        \bottomrule
    \end{tabular}
    \caption{The evolution of $\grade(s)$ and $\mu(s)$ during an example run of Algorithm \ref{alg:necklace}.}
    \label{fig:progress}
\end{figure}

The following lemma explains the relevance of $\mu$.
\begin{restatable}{lemma}{lemmuprogress}
\label{lem:mu-progress}
The progress measure $\mu(s)$ increases
w.r.t.~the lexicographic ordering $<_{lex}$ each time one of the
updates presented in Figures \ref{tab:first}, \ref{tab:umn},
\ref{tab:mn}, \ref{tab:qm}, and \ref{tab:last} takes place.  
\end{restatable}

As in the case of Lemma \ref{lem:grades}, the proof is lengthy and proceeds by a detailed case analysis. It can be found in the appendix.
We are now in position to prove the appropriate result concerning open
chains of cycles.

\begin{theorem}
\label{thm:necklace-noweight-nobonus}
Every coordination game on an open chain of $m$ cycles, each with $v$ nodes, 
ensures improvement paths of length $\leq 3\lcnode m^3$.
\end{theorem}
\proof{\sc Proof.} 
  Let $s_0$ be an arbitrary initial joint strategy in this
  coordination game.  We argue that starting at $s_0$, Algorithm \ref{alg:necklace}
  computes a finite improvement path $\rho$ of length at most
  $3\lcnode m^3$.  By Lemma \ref{lem:mu-progress} $\mu(s)$ increases according to the
  lexicographic order each time the joint strategy $s$ is
  modified in line 8.

We now estimate the number of different values the progress
measure $\mu$ can take.  If $s$ is a Nash equilibrium, then $\mu(s) = \mytop$, which
accounts for one value.
Otherwise $\guard(s) \in \{0,\ldots,m-1\}$ and
$\guard(s)+1 \leq \nbr(s) \leq |\prefix(s)| \leq m,$
because by definition the index $\nbr(s)$ cannot be smaller than
$\guard(s)+1$ and the grade of the cycle with this index
belongs to $\prefix(s)$.  Therefore the number of values
$\mu$ can take is
$$1 + \sum_{g = 0}^{m-1} \sum_{p=g+1}^{m} (p-g) + \sum_{g=0}^{m-1} (m-g) = 
1 + \sum_{g=0}^{m-1} \frac{(m-g)(1+m-g)}{2} + \frac{m(1+m)}{2} = $$
$$1+ \sum_{x=1}^{m} \frac{x(1+x)}{2} + \frac{m(1+m)}{2} = 1 + \frac{m(m+1)(m+2)}{6} + \frac{m(1+m)}{2} = $$

$$1 + \frac{m(m+1)(m+5)}{6} \leq m^3 \text{\ \ for $m \geq 2$.}$$
As a result, the length of the improvement path constructed by
Algorithm \ref{alg:necklace} is at most $3\lcnode m^3$, because by
Theorems~\ref{thm:TARK15a} and Theorem \ref{thm:cycle-2bonuses} the
improvement path in line 6 takes at most $3\lcnode$ improvement steps.
\qed
\endproof

Finally, so far we assumed that we know the decomposition of the game
graph into a chain of cycle in advance.  In general the input may be
an arbitrary graph and we would need to find this decomposition first.
Fortunately this can be done in linear time as the following result shows.

\begin{proposition}
\label{pro:detect-chains}
Checking whether a given graph $G$ is an \ochain{}, and if so
partitioning $G$ into simple cycles
$\mathcal{C}_1, \ldots, \mathcal{C}_m$ can be done in $\calO(|G|)$
time.
\end{proposition}

\proof{\sc Proof.}
  First note that if $G$ is an \ochain{} then there are no bidirectional
  edges and each of its nodes has either out- and in-degree values
  both equal to 1 or both equal to 2.  These two conditions can be easily
  checked in linear time by simply going through all the nodes and
  their edges in $G$.

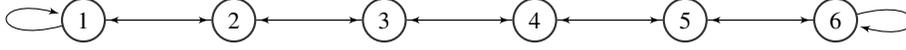
\begin{figure}[htbp]
    \centering
    \tikzstyle{agent}=[circle,draw=black!80,thick, minimum size=1.8em,scale=0.9]
    \begin{tikzpicture}[auto,>=latex',shorten >=1pt,on grid]
    
    \foreach \x in {1,2,3,4,5,6}{
        \draw (2*\x,0) node[agent] (\x) {\x};
    }    
    \foreach \x [evaluate=\x as \y using int(\x+1)] in {1,2,3,4,5}{
        \draw [->] (\x) to (\y);
    }
    \foreach \x [evaluate=\x as \y using int(\x-1)] in {2,3,4,5,6}{
    \draw [->] (\x) to (\y);
}
    \draw[->,loop left,min distance=10mm] (1) to (1);
    \draw[->,loop right,min distance=10mm] (6) to (6);
        
    \end{tikzpicture}
    \caption{Graph $G'$ corresponding to an open chain $G$ with 7 cycles (and 6 link nodes).}
    \label{fig:gprime}
\end{figure}

  Assume that the above two conditions hold. Let $A$ be the set of all nodes in $G$ that we already identified to have out- and in-degrees both equal to 2.
  We first build a new directed graph $G'$ whose set of nodes is $A$
  and there is an edge from $u \in A$ to $v \in A$ iff $v$ is
  reachable from $u$ by traversing only nodes with out- and in-degree
  both equal to 1. We illustrate this construction in Figure \ref{fig:gprime}.

  Such a graph can be built using a single run of the depth first
  search algorithm starting from any node in $A$.  Now note that the
  original graph $G$ is an \ochain{} iff this graph $G'$ is a simple
  path whose two ends have a self-loop and all edges are
  bidirectional.

  This condition can also be checked in linear time, by simply
  following all edges of $G'$ in one direction.  To partition $G$ into
  simple cycles we label one of the end nodes of $G'$ as
  $\nd{1}{1}$. Its only adjacent node we label as $\nd{2}{1}$, the
  other adjacent node of $\nd{2}{1}$ as $\nd{3}{1}$, and so on until
  the node at the other end is of $G'$ labeled as $\nd{m-1}{1}$. These
  are the labels of the link nodes.  The labels of the remaining nodes
  in each cycle $C_j$ for $j \in \{1, \LL, m\}$ can then be simply
  inferred by following the edges in the original graph $G$.
\qed
\endproof

\section{Strong equilibria}
\label{sec:strong}

In this section we study the existence of strong equilibria
and the existence of finite c-improvements paths. 
To start with, we establish two results about the games that have
the strongest possible property, the c-FIP.

First we establish a structural property of a coalitional deviation
from a \NE in our coordination games.  It will be used to prove c-weak
acyclicity for a class of games on the basis of their weak acyclicity.
Note that such a result cannot hold for all classes of graphs because
there exists a coordination game on an undirected graph which is
weakly acyclic but has no strong equilibrium (see \cite{AKRSS16}).

\begin{lemma}
\label{lem:unicolor-cycle}
Consider a coordination game.  Any node involved in a profitable
coalitional deviation from a \NE belongs to a directed simple cycle
that deviated to the same colour.
\end{lemma}
\proof{\sc Proof.}
  Suppose that $s'$ is profitable deviation of a coalition $K$ from a
  \NE $s$.  It suffices to show that each node in $K$ has a neighbour
  in $K$ deviating to the same colour.  Assume that for some player
  $i\in K$ it is not the case.  Then
  \[
    \begin{array}{l}
      \: \phantom{ = } p_i(s) < p_i(s'_K,s_{-K}) \\
      = \sum_{j\in N_j \cap K: s'_{j} = s'_i} w_{j\to i}$ $+ \sum_{j\in N_j \setminus K: s_{j} = s'_i} w_{j\to i}$ $+ \beta(i,s'_i) \\
\leq 0 + \sum_{j \in N_i: s_{j} = s'_i} w_{j\to i} + \beta(i,s'_i) = p_i(s'_i,s_{-i}),
    \end{array}
\]
which contradicts the fact that $s$ is a
\NE.
\qed
\endproof

\begin{theorem} \label{thm:DAG} Every coordination game with bonuses
  on a weighted directed acyclic graph (DAG) has the c-FIP and a
  fortiori a strong equilibrium.  Further, every Nash equilibrium is a
  strong equilibrium.  Finally, the game ensures both improvement
  paths and c-improvement paths of length $\leq n-1$, where ---recall--- $n$ is
  the number of nodes.
\end{theorem}

\proof{\sc Proof.}
  Given a weighted DAG $(V,E)$ on $n$ nodes denote these nodes by $1, \LL, n$ in such a
  way that for all $i,j \in \{1, \LL, n\}$
\begin{equation}
 \label{equ:rank}
\mbox{if $i < j$ then $(j \to i) \not \in E$.}
\end{equation}
So if $i < j$ then the payoff of the node $i$ does not depend
on the strategy selected by the node $j$.

Then given a coordination game whose underlying directed graph is the
above weighted DAG and a joint strategy $s$ we abbreviate the sequence
$p_{1}(s), \LL, p_{n}(s)$ to $p(s)$.  We now claim that
$p: S \to \mathbb{R}^n$ is a generalised ordinal c-potential when we
take for the partial ordering $\succ$ on $p(S)$ the lexicographic
ordering $>_{lex}$ on the sequences of reals.

So suppose that some coalition $K$ profitably deviates from the joint
strategy $s$ to $s'$.  Choose the smallest $j \in K$.  Then
$p_{j}(s') > p_{j}(s)$ and by (\ref{equ:rank}) $p_{i}(s') = p_{i}(s)$
for $i < j$.  By the definition of $>_{lex}$ this implies
$p(s') >_{lex} p(s)$, as desired.  Hence the game has the c-FIP.

The second claim is a direct consequence of Lemma \ref{lem:unicolor-cycle}
that implies that no coalition deviations are possible from a Nash equilibrium for DAGs.

Finally, to prove the last claim,
given an initial joint strategy schedule the players in the
order $1, \LL, n$ and repeatedly update the strategy of each selected
player to a best response. By (\ref{equ:rank}) this yields an
improvement path of length $\leq n-1$. By the second claim this path
is also a c-improvement path.  
\qed
\endproof

Example \ref{exa:rotate} shows that it is difficult to come up with
other classes of directed graphs for which the coordination game has
the FIP, let alone the c-FIP.  However, the weaker property of c-weak
acyclicity holds for the games on simple cycles considered in
Section~\ref{sec:simple-cycle}.
Below we put $i \ominus 1 = i-1$ if $i > 1$ and $1 \ominus 1 = n$.  

\begin{theorem}
\label{thm:se-cycle}
Consider a coordination game with bonuses on a weighted simple cycle.
Any finite improvement path is a finite c-improvement path or can be
extended to it by a single profitable deviation of all players.
\end{theorem}

\proof{\sc Proof.}
  Take a finite improvement path and denote by $s$ the \NE it reaches.
  If $s$ is a strong equilibrium then we are done.  Otherwise there
  exists a coalition $K$ with a profitable deviation from $s$.  By
  Lemma \ref{lem:unicolor-cycle} the coalition $K$ consists of all
  players and all of them switch to the same colour.

  Let $C$ be the set of common colours $c$ such that a switching by
  all players to $c$ is a profitable deviation from $s$.
  We just showed that $C$ is non-empty.
  Select an arbitrary player $i_0$ and choose a colour from $C$ for
  which player $i_0$ has a maximal bonus.  Let $s'$ be the resulting
  joint strategy.

  We first claim that $s'$ is a \NE. Otherwise some player $i$ can
  profitably deviate from $s'_i$ to a colour $c$. Then we have
  $s'_{i \ominus 1} \neq c$, because all players hold the same colour
  in $s'$.  So we have
  $p_i(s) < p_i(s') < p_i(c,s'_{-i}) = \beta(i,c) \leq p_i(c,s_{-i})$,
  which is a contradiction since $s$ is a \NE.

  Next, we claim that $s'$ is a strong equilibrium. Otherwise by the
  initial observation there is a profitable deviation of all players
  from $s'$ to some joint strategy $s''$ in which all players switch
  to the same colour.  So $p_{i_0}(s') < p_{i_0}(s'')$. Moreover, this
  profitable deviation is also a profitable deviation of all players
  from $s$, which contradicts the choice of $i_0$.
\qed
\endproof
\II

The above result directly leads to the following conclusions.

\begin{corollary} \label{cor:TARK15} 

\mbox{}

\begin{enumerate}[(i)]

\item Every coordination game on a weighted simple cycle in which at most
one node has bonuses ensures c-improvement paths of length $\leq 2n$.

\item Every coordination game with bonuses on a simple cycle 
in which at most one edge has a non-trivial weight
ensures c-improvement paths of length $\leq 3n$.

\item Every coordination game on a weighted simple cycle in which 
two nodes have bonuses ensures c-improvement paths of length $\leq 3n+1$.

\item Every coordination game on a simple cycle with bonuses in which
  two edges have non-trivial weights ensures c-improvement
  paths of length $\leq 4n$.

\end{enumerate}
\end{corollary}
\proof{\sc Proof.}
  By Theorems \ref{thm:TARK15a}, \ref{thm:TARK15},
  \ref{thm:cycle-2bonuses}, \ref{thm:cycle-2weights}, and
  \ref{thm:se-cycle}.
\qed
\endproof
\II

We conclude this analysis of coordination games on simple cycles by
the following observation that sheds light on Theorem
\ref{thm:se-cycle} and is of independent interest.

\begin{proposition} \label{pro:k}
  Consider a coordination game with bonuses on a simple
  cycle with $n$ nodes. Then every Nash equilibrium is an
  $(n-1)$-equilibrium.
\end{proposition}
\proof{\sc Proof.}
  Take a Nash equilibrium $s$. It suffices to prove that it is an
  $(n-1)$-equilibrium.  Suppose otherwise. Then for some coalition $K$
  of size $\leq n-1$ and a joint strategy $s'$, $s \betredge{K} s'$ is
  a profitable deviation.

  Take some $i \in K$ such that $i \ominus 1 \not \in K$.  We have
  $p_i(s') > p_i(s)$. Also $p_i(s'_i, s_{-i}) = p_i(s')$, since
  $s_{i \ominus 1} = s'_{i \ominus 1}$.  So
  $p_i(s'_i, s_{-i}) > p_i(s)$, which contradicts the fact that $s$ is
  a Nash equilibrium.
\qed
\endproof

From the definition of an $(n-1)$-equilibrium and
  Proposition~\ref{pro:k}, it follows that for a coordination game with bonuses on
  a simple cycle with $n$ nodes, every Nash equilibrium is a
  $k$-equilibrium for all $k \in \{1, \LL, n-1\}$.  We now show that,
as in the case of simple cycles, coordination games on \ochains are
c-weakly acyclic, so {\it a fortiori} have strong equilibria.

We begin with the following useful fact.

\begin{lemma}\label{lem:unicolor}

  Suppose that in a joint strategy $s$ for the coordination game on an
  open chain of $m$ simple cycles $\mathcal{C}_j$, where
  $j \in \{1, \ldots, m\}$, a simple cycle $\mathcal{C}_i$ is
  unicoloured.  Then in any profitable deviation from $s$ the colours
  of the nodes in $\mathcal{C}_i$ do not change.
\end{lemma}
\proof{\sc Proof.}
The payoff of each node of the cycle $\mathcal{C}_i$
in $s$ is $\geq 1$.  For the non-link nodes the payoff is then
maximal, so none of these nodes can be a member of a coalition that
profitably deviates. This implies that a link node cannot be a member
of a coalition that profitably deviates either. Indeed, otherwise its
payoff increases to 2 and hence in the new joint strategy its colour
is the same as the colour of its predecessor $j$ in the cycle
$\mathcal{C}_i$, which is not the case, since we just explained that
the colour of $j$ does not change.
\qed
\endproof

\begin{theorem}
\label{thm:se-open-chain}
Every coordination game on an open chain of $m$ simple cycles, each with $v$ nodes,
ensures c-improvement paths of length $4vm^4$.
\end{theorem}

\proof{\sc Proof.}
Assume the considered \ochain{} $\mathcal{C}$
consists of the simple cycles
$\mathcal{C}_j$, where $j \in \{1, \ldots, m\}$.  

We now construct the desired c-improvement path $\xi$ as an alternation of
an improvement path guaranteed by Theorem
\ref{thm:necklace-noweight-nobonus} and a single profitable deviation
by a coalition. Each time such a profitable coalitional deviation
takes place, by Lemma \ref{lem:unicolor-cycle} the deviating
coalition includes a simple cycle $\mathcal{C}_i$
all nodes of which switch to the same colour. By Lemma~\ref{lem:unicolor}
each time this is a different cycle,
which is moreover disjoint from the previous cycles.
This implies that the number of such profitable deviations in $\xi$
is at most $\lceil m/2 \rceil$.

So $\xi$ is finite and by Theorem \ref{thm:necklace-noweight-nobonus}
its length is at most
$(\lceil m/2 \rceil + 1) \cdot 3vm^3 + \lceil m/2 \rceil$, where the
first term counts the total length of at most $\lceil m/2 \rceil + 1$
improvement paths that separate at most $\lceil m/2 \rceil$ coalitional
deviations, which is the second term of this expression. But
$\lceil m/2 \rceil + 1 \leq m$ for $m \geq 2$, so
$(\lceil m/2 \rceil + 1 )\cdot 3vm^3 + \lceil m/2 \rceil \leq 3vm^4 +
\lceil m/2 \rceil \leq 4vm^4$.
\qed
\endproof

Example \ref{exa:rotate} shows that even when only two colours are
used, coordination games need not have the FIP.
This is in contrast to the case of undirected
graphs for which we proved in \cite{AKRSS16} that the corresponding class 
of coordination game does have the FIP.  On the other hand, a weaker property
does hold.

\begin{theorem} \label{thm:two-c1}
Every coordination game in which only two colours are used 
ensures improvement paths of length $\leq 2n$.
%
\end{theorem}

\proof{\sc Proof.}
  We prove the result for a more general class of games, namely the
  ones that satisfy the PPM (the property defined in Section
  \ref{sec:prelim}).  Call the colours blue and red. When a node holds
  the blue colour we refer to it as a blue node, and the likewise for the
  red colour.  Take a joint strategy $s$.  \II

\NI
\emph{Phase 1.}
We consider a maximal sequence $\xi$ of profitable deviations starting
in $s$ in which each node can only switch to blue. At each step the
number of blue nodes increases, so $\xi$ is of length at most $n$.
Let $s^1$ be the last joint strategy in $\xi$.  If $s^1$ is a \NE,
then $\xi$ is the desired finite improvement path. Otherwise we move
to the next phase.  \II

\NI
\emph{Phase 2.} We consider a maximal sequence $\chi$ of
profitable deviations starting in $s^1$ in which each
node can only switch to red. Also $\chi$ is of length at most $n$.
Let $s^{2}$ be the last joint strategy in $\chi$.
\II

We claim that $s^{2}$ is a \NE. Suppose otherwise. Then some node, say
$i$, can profitably switch in $s^{2}$ to blue.
Suppose that node $i$ is red in $s^1$.  In $s^1$ there are weakly more blue nodes
than in $s^{2}$, so by the PPM also in $s^{1}$ node $i$ can profitably
switch to blue. This contradicts the choice of $s^{1}$. 

Hence node $i$ is blue in $s^1$, while it is red in $s^2$.  So in some
joint strategy $s^3$ from $\chi$ node $i$ profitably switched to
red.
Then $s^3 = (i:b, s^3_{-i})$ and
$p_i(i:b, s^3_{-i}) < p_i(i:r, s^3_{-i})  \leq p_i(i:r, s^2_{-i}) < p_i(i:b, s^2_{-i}),$
where the weak inequality holds due to the PPM.
But in $s^3$ there are weakly more blue nodes than in $s^2$, so by the PPM
$p_i(i:b, s^{2}_{-i}) \leq  p_i(i:b, s^3_{-i}).$
This yields a contradiction.
\qed
\endproof


The following simple example shows that in the coordination games in
which only two colours are used Nash equilibria do not need to be
strong equilibria.

\begin{example}
  \rm
  
  Consider a bidirectional cycle $1 \lra 2 \lra 3 \lra 4 \lra 1$ in which each node has two colours,
  $a$ and $b$. Then $(a,a,b,b)$ is a Nash equilibrium, but it is not a strong equilibrium
  because of the profitable deviation to $(a,a,a,a)$, which is a strong equilibrium.
\qed
\end{example}

On the other hand the following counterpart of the above result holds
for the c-improvements paths.

\begin{theorem} \label{thm:two-c}
Every coordination game in which only two colours are used 
ensures c-improvement paths of length $\leq 2n$.
%
\end{theorem}

\proof{\sc Proof.}
  As in the above proof we establish the result for the games that
  satisfy the PPM.  We retain the terminology of blue and red colours,
  that we abbreviate to $b$ and $r$.
Take a joint strategy $s$. 
\II

\NI 
\emph{Phase 1.}  
We consider a maximal sequence $\xi$ of profitable deviations of the
coalitions starting in $s$ in which the nodes can only switch to
blue. At each step the number of blue nodes increases, so $\xi$ is of
length at most $n$.  Let $s^1$ be the last joint strategy in $\xi$.
If $s^1$ is a strong equilibrium, then $\xi$ is the desired finite
c-improvement path. Otherwise we move to the next phase.  \II

\NI
\emph{Phase 2.} We consider a maximal sequence $\chi$ of profitable
deviations of the coalitions starting in $s^1$ in which the nodes can
only switch to red. Also $\chi$ is of length at most $n$.  Let $s^{2}$
be the last joint strategy in $\chi$.  \II

We claim that $s^{2}$ is a strong equilibrium. Suppose otherwise.
Then for some joint strategy $s'$, $s^{2} \betredge{K} s'$ is a
profitable deviation of some coalition $K$. Let $L$ be the set of
nodes from $K$ that switched in this deviation to blue.  By the
definition of $s^{2}$ the set $L$ is non-empty.

Given a set of nodes $M$ and a joint strategy $s$ we denote by
$(M:b, s_{-M})$ the joint strategy obtained from $s$ by letting the
nodes in $M$ to select blue, and similarly for the red colour.  Also
it should be clear which joint strategy we denote by $(M:b, P
\setminus M:r, s_{-P})$, where $M \sse P$.
 
We claim that $s^{2} \betredge{L} (L:b, s^{2}_{-L})$ is a
profitable deviation of the players in $L$. Indeed, we have for all
$i \in L$
\begin{equation}
p_i(s^{2}) < p_i(L:b, s^{2}_{-L}), 
  \label{equ:k+l}
\end{equation}
since by the assumption $p_i(s^{2}) < p_i(s')$ and by the PPM
$p_i(s') \leq p_i(L:b, s^{2}_{-L})$.

Let $M$ be the set of nodes from $L$ that are red in $s^1$.
Suppose that $M$ is non-empty. We show that then for all
$i \in M$
\begin{equation}
  \label{equ:M}
  p_i(M:r, L \setminus M:b, s^1_{-L}) < p_i(M:b, L \setminus M:b, s^1_{-L}).
\end{equation}
Indeed, we have for all $i \in M$
\[
\begin{array}{ll}
     & p_i(M:r, L \setminus M:b, s^1_{-L}) \leq p_i(M:r, L \setminus M:b, s^{2}_{-L}) \\
\leq & p_i(M:r, L \setminus M:r, s^{2}_{-L}) < p_i(M:b, L \setminus M:b, s^{2}_{-L}) \\
\leq & p_i(M:b, L \setminus M:b, s^1_{-L}),
\end{array}
\]
where the weak inequalities hold due to the PPM and the
strict inequality holds by the definition of $L$.

But $s^1 = (M:r, L \setminus M:b, s^1_{-L})$, so
(\ref{equ:M}) contradicts the definition of $s^1$.
Thus $M$ is empty, i.e., all nodes from $L$ are blue in $s^1$.

Let $i$ be a node from $L$ that as first turns red in $\chi$.
So in some joint strategy $s^3$ from $\chi$ node $i$
profitably switched to red in a profitable deviation to a joint strategy $s^4$.
Then $s^3 = (L:b, s^3_{-L})$, $s^4 = (i:r, s^4_{-i})$ and
\[
p_i(L:b, s^3_{-L}) < p_i(i:r, s^4_{-i}) \leq p_i(s^2) < p_i(L:b, s^{2}_{-L}), 
\]
where the weak inequality holds due to the PPM and the strict
inequalities hold by the definition of $i$ and (\ref{equ:k+l}).  But
in $(L:b, s^3_{-L})$ there are weakly more
blue nodes than in $(L:b, s^{2}_{-L})$, so by the PPM
$p_i(L:b, s^{2}_{-L}) \leq  p_i(L:b, s^3_{-L}).$
This yields a contradiction.
(The final step in this proof in \cite{ASW16} contained a bug that is
now corrected.)
\qed
\endproof
\II

When the underlying graph is symmetric and the set of strategies for
every node is the same, the existence of strong equilibrium for
coordination games with two colours follows from Proposition 2.2 in
\cite{KBW97a}. Theorem~\ref{thm:two-c} shows a stronger result, namely
that these games are c-weakly acyclic. Example
\ref{exa:1} shows that when three colours are used, Nash
equilibria, so a fortiori strong equilibria do not need to exist.
Finally, note that sometimes strong equilibria exist even though the
coordination game is not c-weakly acyclic.

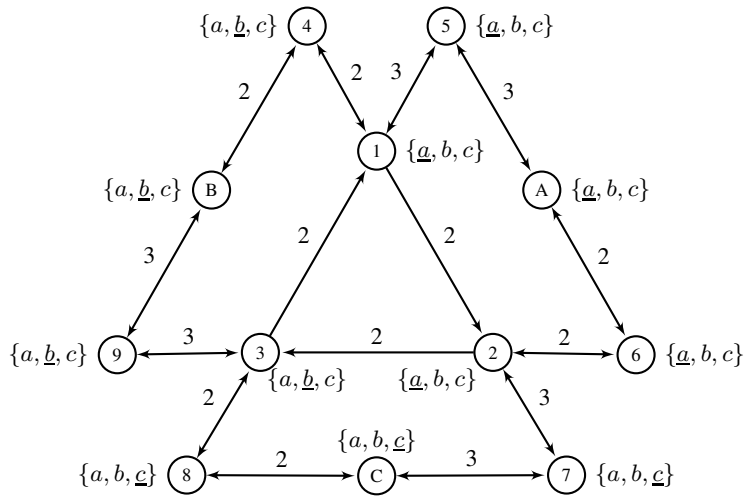
\begin{figure}
\centering
\tikzstyle{agent}=[circle,draw=black,thick, minimum size=2em,scale=0.7]
\tikzstyle{small}=[scale=0.9]
\begin{tikzpicture}[auto,>=latex',shorten >=1pt,on grid,scale=1.05]
\newdimen\R
\R=1.7cm
\newcommand{\llab}[1]{{\small $\{#1\}$}}
\newcommand{\lla}{\llab{\underline{a},b,c}}
\newcommand{\llb}{\llab{a,\underline{b},c}}
\newcommand{\llc}{\llab{a,b,\underline{c}}}
\draw (90: \R) node[agent,label=right:{\lla}] (1) {1};
\draw (90-120: \R) node[agent,label={[label distance=-4pt]below left: {\lla}}] (2) {2};
\draw (90-240: \R) node[agent,label={[label distance=-4pt]below right:\hspace{-1.2mm}{\llb}}] (3) {3};
\draw (90+15: 2*\R) node[agent,label=left:{\llb}] (4) {4};
\draw (90-15: 2*\R) node[agent,label=right:{\lla}] (5) {5};
\draw (-15: 2*\R) node[agent,label=right:{\lla}] (6) {6};
\draw (-15-30: 2*\R) node[agent,label=right:{\llc}] (7) {7};
\draw (180+45: 2*\R) node[agent,label=left:{\llc}] (8) {8};
\draw (180+15: 2*\R) node[agent,label=left:{\llb}] (9) {9};
\draw (30: 1.42*\R) node[agent,label=right:{\lla}] (A) {A};
\draw (120+30: 1.42*\R) node[agent,label=left:{\llb}] (B) {B};
\draw (240+30: 1.42*\R) node[agent,label={[label distance=-2pt]above:{\llc}}] (C) {C};
\foreach \x/\y/\w in {1/2/2,3/1/2} { 
    \draw[->, thick] (\x) to node[small] {\w} (\y) ;
}
\draw[->,thick] (2) to node[small,above] {2} (3);

\foreach \x/\y/\w in {4/1/2,1/5/3,2/6/2,2/7/3,8/3/2,9/3/3} { 
    \draw[<->, thick] (\x) to node[small] {\w} (\y) ;
}
\foreach \x/\y/\w in {5/A/3,A/6/2,B/4/2,9/B/3,8/C/2,C/7/3} { 
    \draw[<->, thick] (\x) to node[small] {\w} (\y) ;
}
\end{tikzpicture}
\caption{\label{fig:no-way-to-se}
A coordination game with strong equilibria unreachable from a given initial joint strategy. 
}
\vskip-0.8em
\end{figure}

\begin{example}
\label{ex:no-way-to-se}
\rm 
Consider the coordination game depicted in Figure \ref{fig:no-way-to-se}.
Note that the underlying graph is strongly connected and that all
edges except $1 \to 2, \: 2 \to 3$ and $3 \to 1$ are bidirectional.
Although the graph is weighted, the weighted edges can be
replaced by unweighted ones by adding auxiliary nodes without
affecting the strong connectedness of the graph.  The behaviour of the
game on this new unweighted graph will be analogous to the one
considered.

Let us analyse the initial joint strategy $s$ that consists of the
underlined colours in Figure \ref{fig:no-way-to-se}.  We argue that
the only nodes that can profitably switch colours (possibly in a
coalition) are the nodes 1, 2 and 3 and that this is the case
independently of their strategies.

First consider the nodes A, B, and C. They have the maximum possible
payoff of 5, independently of the strategies of the nodes 1, 2 and 3,
so none of them can be a member of a profitably deviating coalition.

Further, each node from the set $\{4, \LL, 9\}$ has two neighbours,
each with the same weight. One of them is from the set $\{$A, B, C$\}$
with whom it shares the same colour, which results in the payoff of 2.
So for each node from $\{4, \LL, 9\}$ a possible profitable
coalitional deviation has to involve a neighbour from $\{$A, B, C$\}$.

Therefore, the only nodes that can profitably deviate are nodes 1, 2
and 3.  Moreover, this will continue to be the case in any joint
strategy resulting from a sequence of profitable coalitional
deviations starting from $s$.  (Another way to look at it by arguing
that the restriction of $s$ to the nodes $\{$A, B, C, $4, \LL, 9\}$ is
a strong equilibrium in the game on these nodes in which we add to the
nodes from $\{4, 6, 8\}$ bonuses 2 and to the nodes from $\{5, 7, 9\}$
bonuses 3.)

So it suffices to analyse the weighted simple cycle and the colour
assignment depicted in Figure~\ref{fig:noNE1}, with the non-trivial
bonuses mentioned above the colours.

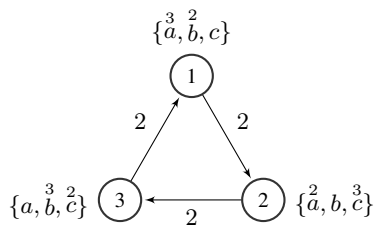
\begin{figure}[htbp]
\normalsize

\centering

\tikzstyle{agent}=[circle,draw=black!80,thick, minimum size=2em,scale=0.8]
\begin{tikzpicture}[auto,>=latex',shorten >=1pt,on grid]
\newdimen\R
\R=1.1cm
\newcommand{\llab}[1]{{\small $\{#1\}$}}
\draw (90: \R) node[agent,label=above:{\llab{\stackrel{3}{a}, \stackrel{2}{b}, c}}] (1) {1};
\draw (90-120: \R) node[agent,label=right:{\llab{\stackrel{2}{a}, b, \stackrel{3}{c}}}] (2) {2};
\draw (90-240: \R) node[agent,label=left:{\llab{a, \stackrel{3}{b},\stackrel{2}{c}}}] (3) {3};
\foreach \x/\y in {1/2,2/3,3/1}{
    \draw[->] (\x) to node {{\small $2$}} (\y); 
}   
\end{tikzpicture}
\caption{A coordination game without a Nash equilibrium \label{fig:noNE1}}
\end{figure}

However, the resulting coordination game does not have a Nash
equilibrium and a fortiori no strong equilibrium. To see it first
notice that each of the nodes can secure a payoff at least 3, while
selecting a colour with a trivial bonus it can secure a payoff of at
most 2. So we do not need to analyse joint strategies in which a node
selects a colour with a trivial bonus. This leaves use with the
following list of joint strategies: $(\underline{a},a,b)$,
$(a,a,\underline{c})$, $(a,c,\underline{b})$, $(a,\underline{c},c)$,
$(b,\underline{a},b)$, $(\underline{b},a,c)$, $(b,c,\underline{b})$
and $(\underline{b},c,c)$.  In each of them, as in Examples
\ref{exa:1} and \ref{exa:simple_cycle}, we underlined a strategy that
is not a best response to the choice of other players.  This means
that no c-improvement path in this game terminates.

Consequently
no c-improvement path in the original game that starts with $s$ terminates.
Therefore, the original game is neither weakly acyclic nor c-weakly acyclic.
On the other hand, it has three trivial strong equilibria in which all
players pick the same colour.  \qed
\end{example}

Note that in the game considered in this example all players have the
same sets of strategies.  We can summarise this example informally as
follows. There exists a graph with the same set of alternatives
(called colours) for all nodes and an initial situation (modelled by a
colour assignment) starting from which no stable outcome (modelled as
a Nash equlibrium) can be achieved even if forming coalitions is
allowed.

\section{Complexity issues}
\label{sec:directed-complexity}

Finally, we study the complexity of finding Nash equilibria and strong
equilibria, and of determining their existence.  The results obtained
so far provide bounds on the length of short (c-)improvement
paths. But in each proof we actually provide bounds on the length of
the corresponding schedule, a notion defined in Section
\ref{sec:prelim}.  This allows us to determine in each case the
complexity of finding a Nash equilibrium or a strong equilibrium,
by analysing the cost of finding a profitable deviation from a given joint strategy.
For the case of weighted graphs we assume all weights to be natural numbers.

We assume that the colour assignment $C$ is given as a $\{0,1\}$-matrix
of size $V \times M$, such that $(i,c)$ entry is $1$ iff colour $c$ is
available to node $i$.  The bonus function $\beta$, if present, is
represented by another matrix of size $V \times M$, where the $(i,c)$
entry holds the value of $\beta(i,c)$.  The game graph is represented
using adjacency lists, where for each node we keep a list of all
outgoing and incoming edges and, if the graph is weighted, their
weights are represented in binary.  As usual, we provide the time
complexity in terms of the number of arithmetic operations performed.
All our algorithms operate only on numbers that are linear in the size
of the input, so the actual number of bit operations is at most
polylogarithmically higher.

Below, as in Table \ref{fig:summary} in Section \ref{sec:intro}, $n$ is
the number of nodes, $|E|$ the number of edges, $l$ the number of
colours, and in the case of the open chains of cycles
$m$ the number of simple cycles in a chain and $v$ the number of nodes in each cycle.
We first determine complexity of finding a best response.

\begin{lemma}
  \label{lem:single-step}
  Consider a coordination game.  Given a joint strategy a best
  response for a player $i$ can be computed in time
  $\mathcal{O}(l + e_i)$, where $e_i$ is the number of incoming edges
  to node $i$.
\end{lemma}
\proof{\sc Proof.}
  We first calculate for each colour the sum of the weights on all
  edges from neighbors of player $i$ with that colour.  This can be
  done by simply iterating over all $e_i$ incoming edges.  We then
  iterate over all of these $l$ values to select any colour with the
  highest such a value.
\qed
\endproof
\II

When we only care about the current payoff of player $i$, then there is no need to iterate over all $l$ colours and we get the following.

\begin{lemma}
    \label{lem:single-payoff}
    Consider a coordination game.  Given a joint strategy the payoff
    of player $i$ can be computed in time $\mathcal{O}(1 +e_i)$, where
    $e_i$ is the number of incoming edges to node $i$.
\end{lemma}
\proof{\sc Proof.}
  It suffices to iterate over all $e_i$ incoming edges the sum the
  weights of all edges from neighbors of player $i$ with the same
  colour. The term $1$ is needed to cover the case of nodes with no neighbours.
\qed
\endproof

We can now deal with the complexity of finding a Nash equilibrium and a
strong equilibrium for the coordination games on simple cycles that we
considered in Section \ref{sec:simple-cycle}.

\begin{theorem} \label{thm:cycles-complexity} Consider a coordination
  game on a simple cycle that is either weighted with at most two
  nodes with bonuses or with bonuses with at most two edges having
  non-trivial weights.  Both a Nash equilibrium and a strong
  equilibrium can be computed in time $\mathcal{O}(nl)$.
\end{theorem}
\proof{\sc Proof.}
  In both cases, due to Theorems \ref{thm:TARK15a}, \ref{thm:TARK15},
  \ref{thm:cycle-2bonuses}, and \ref{thm:cycle-2weights}, to compute a
  Nash equilibrium it suffices to follow a schedule of length
  $\mathcal{O}(n)$.  At each step of this schedule it suffices to
  consider only the deviations to a colour with the maximal bonus. We
  can find such colours in time $\mathcal{O}(l)$ and then simply
  follow the $\mathcal{O}(l)$ procedure given in Lemma
  \ref{lem:single-step} for finding a best response within this
  narrowed down set. We conclude that computing a Nash equilibrium can
  be done in time $\mathcal{O}(nl)$.
	
  Finally, to compute a \SE we first compute a \NE and subsequently
  check whether there is a profitable deviation of all nodes to a
  single colour.  By Theorem \ref{thm:se-cycle} one of these two joint
  strategies is a strong equilibrium.
	
  The latter step involves iterating over all $l$ colours and
  computing for each of them the payoff of all nodes when they all
  hold this single colour, assuming such a colour is shared by all
  nodes. Each iteration takes $\mathcal{O}(n)$ time, which results in
  total $\mathcal{O}(nl)$ time, as well.
\qed
\endproof

The complexity of computing a Nash equilibrium for the coordination games on an open chain of cycles can be easily established as most of the work was done in the proof of
Theorem~\ref{thm:necklace-noweight-nobonus}, that in turn built upon
Theorems \ref{thm:TARK15a} and \ref{thm:cycle-2bonuses}.

\begin{theorem} \label{thm:openchainNE-complexity}
  Consider a coordination game on an open chain of cycles.
 A Nash equilibrium can be computed in time
 $\mathcal{O}(vm^3l)$.

\end{theorem}
\proof{\sc Proof.}
  From Theorem~\ref{thm:necklace-noweight-nobonus} it follows that for
  an open chain of cycles there exists an improvement path of length at most
  $3vm^3$. 
  Due to Lemma~\ref{lem:single-step} computing each best response
can be done in time $\mathcal{O}(l)$. It follows that
  a Nash equilibrium can be computed in time $\mathcal{O}(vm^3l)$.
\qed
\endproof

To analyse the complexity of computing a strong equilibrium for the coordination games on an open chain of cycles we make use of Algorithm \ref{alg:max}.

\begin{algorithm}[ht]
    \caption{\label{alg:max}}
    \SetNlSkip{1em}
    \KwIn{A strategic game $(S_1, \ldots, S_n, p_1, \ldots, p_n)$ that
      satisfies the PPM property, a joint strategy $s$, and a strategy
      $c$.}
    \KwOut{A maximal coalition that can profitably deviate to
      $c$, if there exists one, and otherwise the empty set.}

$A := \{i \in \{1, \LL, n\} \mid c \in S_i\}$; \hskip2em {\it (i.e., $A$ is the set of players that can select $c$)}

\While{\rm $A \neq \ES$ and $s \betredge{A} s'$, where $s'_i = c$ for $i \in A$, is not a profitable deviation}
{choose some $a \in A$ such that  $p_a(s) \geq p_a(s')$;

$A := A \setminus \{a\};$}  

{\bf return} $A$

\end{algorithm}

The following lemma establishes the correctness of Algorithm \ref{alg:max}.

\begin{lemma} \label{lem:max}
  Consider a strategic game that satisfies the PPM property, a joint strategy
  $s$ and a strategy $c$. Algorithm \ref{alg:max} computes a maximal
  coalition that can profitably deviate from $s$ to $c$, if there
  exists one, and otherwise returns the empty set.
\end{lemma}

\proof{\sc Proof.}
First note that due to line 4 the algorithm always terminates.  
Suppose that $A^*$ is a maximal coalition that
can profitably deviate to $c$. So $s \betredge{A^*} s^*$, where
$s^*_i = c$ for $i \in A^*$.  Consider the execution of the above
algorithm. Then $A^* \subseteq A$ after line 1. By
the PPM property no player from $A^*$ can be
removed in line 4, because otherwise it could not profit from the
deviation $s \betredge{A^*} s^*$ either.  So the coalition $A$ the
algorithm returns contains $A^*$ and a fortiori
is non-empty.  Hence the \textbf{while} loop was exited because
$s \betredge{A} s'$, where $s'_i = c$ for $i \in A$, is a profitable
deviation.  By the maximality of $A^*$ we get $A = A^*$.

If no coalition can profitably deviate to $c$, then the
\textbf{while} loop is exited because $A = \ES$ and the algorithm
returns the empty set.
\qed
\endproof
\II

This lemma and Theorem~\ref{thm:se-open-chain}
allow us to derive the following result.

\begin{theorem} \label{thm:openchainSE-complexity} Consider a
  coordination game on an open chain of cycles. A strong
  equilibrium can be computed in time $\mathcal{O}(vm^4l)$.
\end{theorem}
\proof{\sc Proof.}
  By Theorem \ref{thm:se-open-chain} it follows that for an open chain
  of cycles there exists a $c$-improvement path of length at most
  $4vm^4$.  Moverover, such a path consists of $\mathcal{O}(vm^4)$
  single-player improvement steps and $\mathcal{O}(m)$ of
  c-improvement steps.  By Lemma \ref{lem:single-step}, executing
  the former steps can be done in time $\mathcal{O}(vm^4l)$.  It
  remains to estimate the latter.
  
  All considered c-improvement steps are from a \NE. So by Lemma
  \ref{lem:unicolor-cycle}, any node involved in a c-improvement step
  belongs to a directed simple cycle that deviated to the same colour.
  It follows that
  in any c-improvement step, nodes that deviate to two different
  colours cannot be adjacent to each other and so do not influence
  each other payoffs. Therefore, any multicolour c-improvement step
  can be split into a sequence of unicolour c-improvement steps (one
  for each deviating colour).

Consider now a Nash equilibrium $s$ that is not a strong equilibrium.
Each coordination game satisfies the PPM property, so Lemma
\ref{lem:max} implies that by executing Algorithm \ref{alg:max} for each
colour $c$ in turn we eventually find a maximal coalition that can
profitably deviate from $s$ to the same colour or determine that no such coalition exists.

Let us now estimate the time complexity of executing Algorithm
\ref{alg:max}.  Executing the assignment in line 1 can be done in
$\mathcal{O}(vm)$ time. Computing the payoffs of every node in $s$ and
$s'$ in line 2 can be done in $\mathcal{O}(vm)$ time due to Lemma
\ref{lem:single-payoff}.  The \textbf{while} loop can be reentered at
most $vm$ times, because there are at most $vm$ nodes in $A$.
Further, because we are dealing with an open chain of cycles each
removal of a node from $A$ affects the payoff of at most two other
players. So updating the payoffs of all players in $s'$ can be done in
$\mathcal{O}(1)$ time.  Therefore executing the \textbf{while} loop
takes in total $\mathcal{O}(vm)$ time. This is also the time
complexity of executing the algorithm, since line 5 takes only
$\mathcal{O}(1)$ time.
  
To find a unicolour profitable deviation from a \NE that is not a
strong equilibrium, in the worst case Algorithm \ref{alg:max} has to
be executed for each colour. So each such c-improvement step takes in
total $\mathcal{O}(vml)$ time. As there are $\mathcal{O}(m)$ of these
c-improvement steps, their execution takes in total
$\mathcal{O}(vm^2l)$ time. So the execution of these steps is
dominated by the executions of the already considered single-player
improvement steps that take in total $\mathcal{O}(vm^4l)$ time, which
is then also the time bound for computing a strong equilibrium.
\qed
\endproof
\II

Finally, we deal with the cases of weighted DAGs and games with two colours.

\begin{theorem} \label{thm:DAGs-complexity}
Consider a coordination game on a weighted DAG. Both a Nash
equilibrium and a strong equilibrium can be computed in time
$\mathcal{O}(nl + |E|)$.
\end{theorem}

\proof{\sc Proof.}
  Consider a weighted DAG $(V,E)$.  The procedure given in
  Theorem~\ref{thm:DAG} first relabels the nodes using $\{1, \LL, n\}$
  in such a way that for all $i,j \in \{1, \LL, n\}$ if $i < j$, then
  $(j \to i) \not \in E$.  Such a relabelling can be done in time
  $\mathcal{O}(n+|E|)$ by means of a topological sort of nodes using
  a DFS algorithm.  Next, the schedule that we will use is simply
  $1,\ldots,n$. Due to Lemma \ref{lem:single-step}, given a joint
  strategy the best response for a player $i$ can be computed in
  $\mathcal{O}(l + e_i)$ time, where $e_i$ is the number of incoming
  edges to node $i$.

  Thus a Nash equilibrium can be constructed in time
  $\mathcal{O}(\sum_{i\in V} (l + e_i) ) = \mathcal{O}(nl + |E|)$.  By
  Theorem~\ref{thm:DAG} every Nash equilibrium is also a strong
  equilibrium.
\qed
\endproof

\begin{theorem} \label{thm:two-se}
Consider a coordination game on a graph $(V,E)$ in which only two colours
are used.

\begin{enumerate}[(i)]
\item 
A Nash equilibrium can be computed in time $\mathcal{O}(n+|E|)$.

\item 
A strong equilibrium can be computed in time $\mathcal{O}(n^2 + n|E|)$.
\end{enumerate}
\end{theorem}
\proof{\sc Proof.}
Given node $i$ we denote by $e_i$ the number of incoming edges to $i$
and by $e'_i$ the number of outgoing edges from $i$.
\II

\NI $(i)$ The proof of Theorem~\ref{thm:two-c1} provides an algorithm that
follows two phases to construct a Nash equilibrium. In the first phase, 
it constructs a maximal sequence of profitable deviations to the first colour (called blue).
And in the second phase, it does the same for the second colour (called red).
Note that by Lemma \ref{lem:single-payoff}, given a joint strategy,
the payoff of player $i$ can be computed in $\mathcal{O}(1 + e_i)$ time.
Therefore, a profitable deviation from any joint strategy (if it
exists) can be found in time
$\sum_{i \in V} \mathcal{O}(1 + e_i) = \mathcal{O}(n + |E|)$.

This yields time complexity of $\mathcal{O}(n^2 + n|E|)$ for both the
first and the second phase, because each phase consists of at
most $n$ profitable deviations. We can reduce this to $\mathcal{O}(n + |E|)$
by precomputing for every player his payoff for selecting each colour 
and then updating these values as players switch strategies. 
Formally, we proceed as follows.

For each player $i$, given a joint strategy of its opponents, let
$(r_i, b_i)$ be its payoffs for selecting, respectively, red and blue
colours.  By Lemma \ref{lem:single-payoff}, given an initial joint
strategy, these pairs of payoffs for all players can be calculated in
time $\sum_{i \in V} \mathcal{O}(1 + e_i) = \mathcal{O}(n + |E|)$.  
In the first phase, where players switch colour from red to blue only, we simultaneously create a list $L$ of all players $i$ whose current colour is red and $r_i < b_i$ holds.

We then repeatedly remove a player $i$ from $L$ and switch its colour to blue. 
This change affects the payoffs of $e'_i$ other players.
More precisely, $e'_i = |\{j \in V \mid i \in N_j\}|$ and for any $j$
such that $i \in N_j$, the pair $(r_j, b_j)$ is updated to
$(r_j - w_{i\to j}, b_j + w_{j\to i})$.
If after this change $r_j < b_j$ holds and player $j$ holds colour red then we add player $j$ to the list $L$.
Note that no player has be removed from $L$ as a result of the deviation of player $i$ due to the PPM property of our games.
Therefore, after a deviation
of player $i$, the time needed to update all values of $(r_j, b_j)$ and the list $L$ is $\mathcal{O}(1 + e'_i)$.

The first phase ends when $L$ becomes empty. Then we rebuild
the list by switching the role of the colours and proceed in the
analogous way. In particular, from that moment on we add a player $i$
to the list if $b_i < r_i$.

In each phase each player can switch its colour
at most once, so the complexity of each phase, as well as both of
them, is $\sum_{i\in V} \mathcal{O}(1 + e'_i)$ = $\mathcal{O}(n + |E|)$.

\II

\NI
$(ii)$
The existence of c-improvement paths of length at most $2n$ is
guaranteed by Theorem~\ref{thm:two-c}. The algorithm 
follows two phases to construct a strong equilibrium. In the first phase, 
it constructs a maximal sequence, $\xi$, of profitable coalition deviations to the first colour (called blue). And in the second phase, it does the same for the second colour (called red) to construct a sequence $\chi$.
  It now suffices to estimate the time
 complexity of computing a single c-improvement step in the sequences $\xi$ and $\chi$.

In each such step a coalition is selected that deviates profitably to
a single colour, blue or red, the joint strategy is modified, and the
payoffs of the players are appropriately modified.  Without loss of
generality we can assume that each time a maximal coalition is
selected.  By Lemma \ref{lem:max} such a coalition can be computed
using Algorithm \ref{alg:max}.  So it suffices to determine the
complexity of Algorithm \ref{alg:max} and of the computation of the
new joint strategy and the modified payoffs in case of coordination
games with two colours.

The complexity of executing the assignment in line 1 is
$\mathcal{O}(n)$.  To evaluate the condition of the {\bf while} loop
in line 2, we first calculate $p_i(s)$ and $p_i(s')$ for every player
$i$. By Lemma \ref{lem:single-payoff} all these values and the set of
players $A' := \{i \in A \mid p_i(s) \geq p_i(s') \}$, for which the
deviation to $s'$ is not profitable, can be calculated in time
$\sum_{i \in V} \mathcal{O}(1 + e_i) = \mathcal{O}(n + |E|)$. Note that the body
of the {\bf while} loop is executed as long as $A' \neq \emptyset$.

After each removal of a node $a \in A'$ from $A$ in line 4 (and as a
result from $A'$), the payoffs $p_i(s')$ of at most $e_a$ other
players are affected and by Lemma \ref{lem:single-payoff}
updating them takes time
$\mathcal{O}(1+e_a)$. 
At the same time, if for any of these $e_a$ players, the deviation to $s'$ is not longer profitable, i.e., 
$p_i(s) \geq p_i(s')$ holds, then we add him to $A'$.
Note that no player has to be removed from $A'$ after a deviation of player $a$ due to the PPM property of our games.

Now, each player is removed in line 4 at most
once, so the total time needed to execute this {\bf while} loop is
$\sum_{i \in V} \mathcal{O}(1+e'_i) = \mathcal{O}(n+|E|)$ time.
Finally, line 5 takes $\mathcal{O}(1)$ time.  So for both colours the
execution of Algorithm \ref{alg:max} takes $\mathcal{O}(n + |E|)$
time.  Once the algorithm returns the empty set we switch the colours
and move to the second phase.  This phase ends when the algorithm
returns the empty set. By Theorem~\ref{thm:two-c} it follows that a
strong equilibrium can be computed in time $\mathcal{O}(n^2 + n |E|)$.
\qed
\endproof

Finally, we study the complexity of determining the existence of Nash
equilibria and of strong equilibria.  We already noticed in Example
\ref{exa:payoff} that some coordination games have no Nash
equilibria. In general, the following holds.

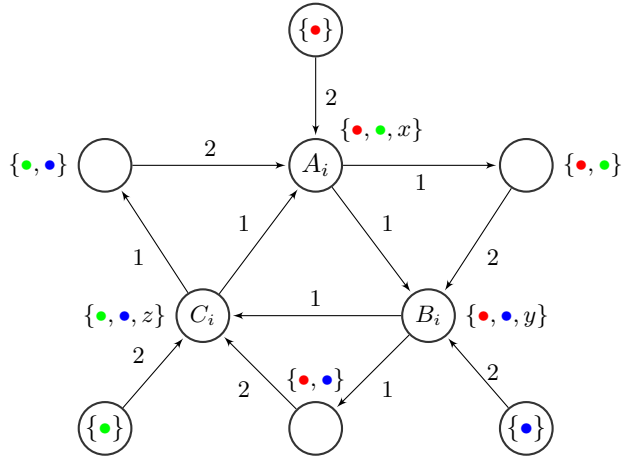
\begin{figure}[htbp]
	\centering
	\newcommand{\redp}[1]{\textcolor{red}{#1}}
	\newcommand{\bluep}[1]{\textcolor{blue}{#1}}
	\newcommand{\greenp}[1]{\textcolor{green}{#1}}
	\tikzstyle{agent}=[circle,draw=black!80,thick, minimum size=0.7cm, inner sep=0pt]
	\tikzstyle{changeblue}=[draw=blue!80,fill=blue!30,thick]
	\tikzstyle{changered}=[draw=red!80,fill=red!30,thick]
	\tikzstyle{changebrown}=[draw=brown!80,fill=brown!40,thick]
	\begin{tikzpicture}[auto,>=latex',shorten >=1pt,on grid]

	\node[agent](aAi) {\small $\{ \redp{\bullet} \}$};
	\node[agent,below of=aAi, node distance=1.8cm,label={[label distance=-.1cm]45:{\small $\{\redp{\bullet},\greenp{\bullet}, x\}$}}](Ai){$A_i$};
	\node[below of=Ai,node distance=2cm,coordinate] (cc){};
	\node[agent,left of=cc, node distance=1.5cm, label=180:{\small  $\{\greenp{\bullet},\bluep{\bullet},z\}$}](Ci){\small $C_i$};
	\node[agent,right of=cc, node distance=1.5cm, label=0:{\small $\{\redp{\bullet},\bluep{\bullet}, y\}$}](Bi){\small $B_i$};
	\node[below of=Ci,node distance=1.5cm,coordinate](bCi){};
	\node[agent, below of=cc,node distance=1.5cm,label=above:{\small $\{\redp{\bullet},\bluep{\bullet}\}$}](bAi){\small $$};
	\node[agent, left of=Ai,node distance=2.8cm,label=180:{\small $\{\greenp{\bullet},\bluep{\bullet}\}$}](lAi){\small $$};
	\node[agent, right of=Ai,node distance=2.8cm,label=0:{\small $\{\redp{\bullet},\greenp{\bullet}\}$}](rAi){\small $$};
	\node[below of=Bi,node distance=1.5cm,coordinate](bBi){};
	\node[agent,left of=bCi,node distance=1.3cm](lbCi){$\{\greenp{\bullet}\}$};
	\node[agent,right of=bBi,node distance=1.3cm](rbBi){$\{\bluep{\bullet}\}$};
	\draw[->] (aAi) to node {{\small $2$}} (Ai);
	\draw[->] (lbCi) to node {{\small $2$}} (Ci);
	\draw[->] (rbBi) to node [right] {{\small $2$}} (Bi);
	\draw[->] (Ai) to node {{\small $1$}} (Bi);
	\draw[->] (Bi) to node [above] {{\small $1$}} (Ci);
	\draw[->] (Ci) to node {{\small $1$}} (Ai);

	\draw[->] (bAi) to node {{\small $2$}} (Ci);
    \draw[->] (Bi) to node {{\small $1$}} (bAi);
	\draw[->] (Ci) to node {{\small $1$}} (lAi);
	\draw[->] (lAi) to node {{\small $2$}} (Ai);
	\draw[->] (Ai) to node[below] {{\small $1$}} (rAi);
	\draw[->] (rAi) to node {{\small $2$}} (Bi);
	\end{tikzpicture}
\caption{Gadget $D_i$ with three parameters $x,y,z \in \{\top,\bot\}$ \label{fig:gadget} and three distinguished nodes $A_i, B_i, C_i$.}
\end{figure}

\begin{theorem} \label{thm:NE-NPcomplete}
The Nash equilibrium existence problem in coordination games without bonuses 
(on unweighted graphs) is NP-complete.
\end{theorem}
\proof{\sc Proof.}
  The problem is in NP, since we can simply guess a colour assignment
  and checking whether it is a Nash equilibrium can be done in polynomial time.
  
  To prove NP-hardness we first provide a reduction from the 3-SAT
  problem, which is NP-complete, to coordination games on directed
  graphs with natural number weights.  Assume we are given a 3-SAT
  formula
$\phi = (a_1 \vee b_1 \vee c_1) \wedge (a_2 \vee b_2 \vee c_2) \wedge \ldots \wedge (a_k \vee b_k \vee c_k)$
with $k$ clauses and $n$ propositional variables $x_1, \ldots, x_n$,
where each $a_i,b_i,c_i$ is a literal equal to $x_j$ or $\lnot x_j$
for some $j$. We will construct a coordination game $\game$ of size
$\mathcal{O}(k)$  with natural number weights 
such that $\game$ has a Nash equilibrium iff $\phi$
is satisfiable.

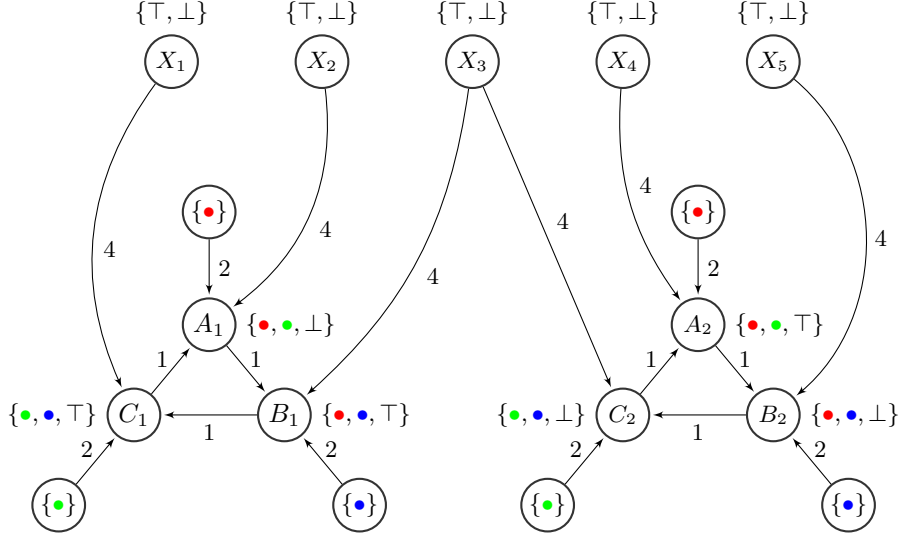
\begin{figure*}
\centering
\newcommand{\redp}[1]{\textcolor{red}{#1}}
\newcommand{\bluep}[1]{\textcolor{blue}{#1}}
\newcommand{\greenp}[1]{\textcolor{green}{#1}}
\centering
\tikzstyle{agent}=[circle,draw=black!80,thick, minimum size=0.7cm, inner sep=0pt]
\tikzstyle{changeblue}=[draw=blue!80,fill=blue!30,thick]
\tikzstyle{changered}=[draw=red!80,fill=red!30,thick]
\tikzstyle{changebrown}=[draw=brown!80,fill=brown!40,thick]
\begin{tikzpicture}[auto,>=latex',shorten >=1pt,on grid]
\begin{scope}
\node[agent,label=above:{\small $\{\top, \bot\}$}](z1){\small{$X_1$}};
\node[agent, right of=z1, node distance=2cm,label=above:{\small $\{\top, \bot\}$}](z2){\small $X_2$};
\node[agent, right of=z2, node distance=2cm,label=above:{\small $\{\top, \bot\}$}](z3){\small $X_3$};
\node[agent, right of=z3, node distance=2cm,label=above:{\small $\{\top, \bot\}$}](z4){\small $X_4$};
\node[agent, right of=z4, node distance=2cm,label=above:{\small $\{\top, \bot\}$}](z6){\small $X_5$};
\end{scope}
\begin{scope}[xshift=0.5cm,yshift=-2cm]
\node[agent](a1) {\small $\{\redp{\bullet}\}$};
\node[agent,below of=a1, node distance=1.5cm,label=right:{\small $\{\redp{\bullet},\greenp{\bullet}, \bot\}$}](b1){$A_1$};
\node[below of=b1,node distance=1.2cm,coordinate] (cc){};
\node[agent,left of=cc, label=180:{\small  $\{\greenp{\bullet},\bluep{\bullet}, \top \}$}](c1){$C_1$};
\node[agent,right of=cc,label=0:{\small $\{\redp{\bullet},\bluep{\bullet}, \top\}$}](c2){$B_1$};
\node[below of=c1,node distance=1.2cm,coordinate](dc1){};
\node[below of=c2,node distance=1.2cm,coordinate](dc2){};
\node[agent,left of=dc1](d1){\small $\{\greenp{\bullet}\}$};
\node[agent,right of=dc2](d2){\small $\{\bluep{\bullet}\}$};
\draw[->] (b1) to node[inner sep=1pt] {{\small $1$}} (c2);
\draw[->] (c2) to node {{\small $1$}} (c1);
\draw[->] (c1) to node[inner sep=1pt] {{\small $1$}} (b1);
\draw[->] (a1) to node {{\small $2$}} (b1);
\draw[->] (d1) to node[inner sep=1pt] {{\small $2$}} (c1);
\draw[->] (d2) to node [inner sep=1pt,swap] {{\small $2$}} (c2);
\draw[->](z1) to [bend right=30] node{{\small $4$}} (c1);
\draw[->](z2) to [bend right=-30] node{{\small $4$}} (b1);
\draw[->](z3) to [bend right=-20] node{{\small $4$}} (c2);
\end{scope}

\begin{scope}[xshift=7cm,yshift=-2cm]
\node[agent,label=right:{}](a1) {\small $\{\redp{\bullet}\}$};
\node[agent,below of=a1, node distance=1.5cm,label=right:{\small $\{\redp{\bullet},\greenp{\bullet},\top\}$}](b1){\small{$A_2$}};
\node[below of=b1,node distance=1.2cm,coordinate] (cc){};
\node[agent,left of=cc,label=180:{\small  $\{\greenp{\bullet},\bluep{\bullet},\bot\}$}](c1){\small{$C_2$}};
\node[agent,right of=cc,label=0:{\small $\{\redp{\bullet},\bluep{\bullet},\bot\}$}](c2){\small{$B_2$}};
\node[below of=c1,node distance=1.2cm,coordinate](dc1){};
\node[below of=c2,node distance=1.2cm,coordinate](dc2){};
\node[agent,left of=dc1,label=right:{}](d1){\small $\{\greenp{\bullet}\}$};
\node[agent,right of=dc2,label=left:{}](d2){\small $\{\bluep{\bullet}\}$};
\draw[->] (b1) to node[inner sep=1pt] {{\small $1$}} (c2);
\draw[->] (c2) to node {{\small $1$}} (c1);
\draw[->] (c1) to node[inner sep=1pt] {{\small $1$}} (b1);
\draw[->] (a1) to node {{\small $2$}} (b1);
\draw[->] (d1) to node[inner sep=1pt] {{\small $2$}} (c1);
\draw[->] (d2) to node [inner sep=1pt,swap] {{\small $2$}} (c2);
\draw[->](z3) to node{{\small $4$}} (c1);
\draw[->](z4) to [bend right=20] node{{\small $4$}} (b1);
\draw[->](z6) to [bend right=-50] node{{\small $4$}} (c2);
\end{scope}

\end{tikzpicture}
\caption{\label{fig:exa}The game $\game$ corresponding to the formula $\phi = (\lnot x_2 \vee x_3 \vee x_1)   \wedge (x_4 \vee \lnot x_5 \vee \lnot x_3)$, where in each gadget the nodes of indegree 1 are omitted.}
\end{figure*}

First, for every propositional variable $x_i$ we have a corresponding
node $X_i$ in $\game$ with two possible colours $\top$ and $\bot$.
Intuitively, for a given truth assignment, if $x_i$ is true then
$\top$ should be chosen for $X_i$ and otherwise $\bot$ should be
chosen.  In our construction we make use of a gadget,
denoted by $D_i(x,y,z)$, with three parameters $x,y, z \in \{\top,
\bot\}$ and $i$ used just for labelling purposes, and presented in
Figure \ref{fig:gadget}.  This gadget behaves similarly to the game
without Nash equilibrium analysed in Example \ref{exa:payoff}.

What is important is that for all possible parameters values, the
gadget $D_i(x,y,z)$ does not have a Nash equilibrium. Indeed, each of
the nodes $A_i$, $B_i$, or $C_i$ can always secure a payoff 2, so
selecting $\top$ or $\bot$ is never a best response and hence in no
Nash equilibrium a node chooses $\top$ or $\bot$. The rest of the
reasoning is as in Example \ref{exa:payoff}.
For any literal $l$, let 
\[
\isPos(l) := \begin{cases}
 \top & \mbox{if $l$ is a positive literal} \\
 \bot & \mathrm{otherwise.}
\end{cases}
\]

For every clause $(a_i \vee b_i \vee c_i)$ in $\phi$ we
add to the game graph $\game$ the $D_i(\isPos(a_i),
\isPos(b_i), \isPos(c_i))$ instance of the gadget.
Finally, for every literal $a_i$, $b_i$, or $c_i$ in $\phi$, which is
equal to $x_j$ or $\lnot x_j$ for some $j$, we add an edge from $X_j$
to $A_i$, $B_i$, or $C_i$, respectively, with weight $4$.
We depict an example game $\game$ in Figure \ref{fig:exa}.
(This Figure corrects the corresponding figure in \cite{ASW16}).
We claim that $\game$ has a Nash equilibrium iff $\phi$ is
satisfiable.
 
\smallskip \noindent ($\Rightarrow$) Assume there is a Nash
equilibrium $s$ in the game $\game$.  We claim that the truth
assignment $\nu :\{x_1, \ldots, x_n\} \to \{\top,\bot\}$ that assigns
to each $x_j$ the colour selected by the node $X_j$ in $s$ makes
$\phi$ true.  Fix $i \in \{1, \LL, k\}$. We need to show that $\nu$
makes one of the literals $a_i$, $b_i$, $c_i$ of the
clause $(a_i \vee b_i \vee c_i)$ true.

From the above observation about the gadgets it follows that at least
one of the nodes $A_i, B_i$, $C_i$ selected in $s$ the same colour as
its neighbour $X_j$.  Without loss of generality suppose it is $A_i$.
The only colour these two nodes, $A_i$ and $X_j$, have in common is
$\isPos(a_i)$. So $X_j$ selected in $s$ $\isPos(a_i)$, which by the
definition of $\nu$ equals $\nu(x_j)$. Moreover, by construction $x_j$
is the variable of the literal $a_i$.  But $\nu(x_j) = \isPos(a_i)$
implies that $\nu$ makes $a_i$ true.

\smallskip \noindent ($\Leftarrow$) Assume $\phi$ is satisfiable. Take
a truth assignment $\nu :\{x_1, \ldots, x_n\} \to \{\top,\bot\}$ that
makes $\phi$ true.  For all $j$, we assign the
colour $\nu(x_j)$ to the node $X_j$. We claim that this assignment can be extended to a
Nash equilibrium in $\game$.

Fix $i \in \{1, \LL, k\}$ and consider the $D_i(\isPos(a_i),
\isPos(b_i), \isPos(c_i))$ instance of the gadget.  The truth
assignment $\nu$ makes the clause $(a_i \vee b_i \vee c_i)$ true.
Suppose without loss of generality that $\nu$ makes $a_i$ true. We
claim that then it is always a unique best response for the node $A_i$
to select the colour $\isPos(a_i)$.

Indeed, let $j$ be such that $a_i = x_j$ or $a_i = \lnot x_j$.  Notice
that the fact that $\nu$ makes $a_i$ true implies that $\nu(x_j) =
\isPos(a_i)$. So when node $A_i$ selects $\isPos(a_i)$, the colour
assigned to $X_j$, its payoff is 4.

This partial assignment of colours can be completed to a Nash
equilibrium. Indeed, remove from the directed graph of $\game$ all
$X_j$ nodes and the nodes that secured the payoff 4, together with the
edges that use any of these nodes. The resulting graph has no cycles,
so by Theorem \ref{thm:DAG} the corresponding coordination game has a
Nash equilibrium.  Combining both assignments of colours we obtain a
Nash equilibrium in $\game$.

To conclude the result for coordination games without weights notice
that an edge with a natural number weight $w$ can be simulated by
adding $w$ extra players to the game.  More precisely, an edge
$(i \to j)$ with the weight $w$ can be simulated by the extra set of
players $\{i_1, \ldots, i_w\}$ and the following $2\cdot w$ unweighted
edges:
$\{(i \to i_1), (i \to i_2), \ldots, (i \to i_w), (i_1 \to j), (i_2
\to j), \ldots, (i_w \to j)\}$.  Given a colour assignment in the
original game with the weighted edges, we then assign to each of the
new nodes $i_1, \LL, i_w$ the colour set of the node $i$.
Then the initial coordination game has a \NE iff the new one, without weights, has one.
Further, the new game can be constructed 
in linear time.
\qed
\endproof

\begin{corollary}
The strong equilibrium existence problem in coordination games
without bonuses (on unweighted graphs) is NP-complete.
\end{corollary}

\proof{\sc Proof.}
  It suffices to note that in the above proof the ($\Rightarrow$)
  implication holds for a strong equilibrium, as well, while in the
  proof of the ($\Leftarrow$) implication by virtue of Theorem
  \ref{thm:DAG} actually a strong equilibrium is constructed.
\qed
\endproof
\II

An interesting application of Theorem \ref{thm:NE-NPcomplete} is in
the context of polymatrix games introduced in Section
\ref{sec:prelim}.  It was shown in \cite{SA15} that deciding whether a
polymatrix game has a Nash equilibrium is NP-complete.  We can
strengthen this result by showing that the problem is strongly
NP-hard, i.e., NP-hard even if all input numbers
are bounded by a polynomial in the size of the input.

\begin{theorem}
\label{thm:polymatrix-NP}
Deciding whether a polymatrix game has a Nash equilibrium is strongly NP-complete.
\end{theorem}

\proof{\sc Proof.}
  Any coordination game $\mathcal{G} = (G,C)$ on an unweighted graph
  $G = (V,E)$ can be viewed as a polymatrix game $\mathcal{P}$ whose
  values of all partial payoffs functions are equal either 0 or 1.
  Specificially, the set of players in $\mathcal{P}$ is the same as in
  $\mathcal{G}$, i.e., $V$. The strategy set $S_i$ of player $i$ is
  simply $C(i)$. We define
  \[
  a^{ij}(\strprofile_i,\strprofile_j) := \begin{cases}
  1 & \mbox{if } j \in N_i \mbox{ and } \strprofile_i = \strprofile_j \\
  0 & \mathrm{otherwise}
  \end{cases}
  \]
  where, as before, $N_i$ is the set of neighbours of node $i$ in the assumed directed graph $G$.
  Notice that the payoffs in both games are the same since for any joint strategy
  $\strprofile=(\strprofile_1,\ldots,\strprofile_n)$,
$
  \payoff^\mathcal{P}_i(\strprofile)=\sum_{j \neq i}
  a^{ij}(\strprofile_i,\strprofile_j) = |\{j \in N_i \mid s_i = s_j\}|
  = p^\mathcal{G}_i(s).
 $
  NP-hardness follows, because this problem was shown to be 
  {\sc NP}-hard for coordination games on unweighted graphs in Theorem
  \ref{thm:NE-NPcomplete}. 
As all numerical inputs are assumed to be 0 or 1 they are obviously
  bounded by a polynomial in the size of the input. So strong
  NP-hardness follows.  As shown in \cite{SA15}, deciding whether a
  given polymatrix game has a Nash equilibrium is in NP, which
  together implies strong NP-completeness of this problem.
\qed
\endproof

\section{Conclusions}
\label{sec:conclusions}
In this paper we studied natural coordination games on weighted
directed graphs, in presence of bonuses representing individual
preferences. In our presentation we focussed on the existence of Nash
and strong equilibria and on ways of computing them efficiently in case
they exist. To this end we extensively used improvement and
coalitional improvement (in short c-improvement) paths that can be
seen as an instance of a local search.

We identified natural classes of graphs for which coordination games
have improvement or c-improvement paths of polynomial length. For
simple cycles these results are optimal in the sense that lifting any
of the imposed restrictions may result in coordination game without a
Nash equilibrium.

In proving our results we used increasingly more complex ways of
constructing (c-)improvement paths of polynomial length. In
particular, the construction in the proof of Theorem
\ref{thm:necklace-noweight-nobonus} relied on the constructions
considered in the proofs of Theorems~\ref{thm:TARK15a} and
\ref{thm:cycle-2bonuses}.

For the class of graphs we considered, local search in the form of the
(c-)improvement paths turns out to be an efficient way of computing a
Nash equilibrium or a strong equilibrium. But this is not true in
general.  In fact, Example \ref{ex:no-way-to-se} shows that this form
of local search does not guarantee that a Nash equilibrium or a strong
equilibrium can be found, even when the underlying graph is strongly
connected and all nodes have the same set of colours.  We also showed
that the existence problem both for Nash and strong equilibria is
NP-complete even for the coordination games on unweighted graphs and
without bonuses.

There are other directed graphs than the ones we considered here, for
which the coordination games are weakly or c-weakly acyclic.  For
example, we proved in \cite{AKRSS16} that the coordination games on
complete graphs have the c-FIP and the proof carries through to the
complete directed graphs. In turn, in \cite{ASW16} we showed that that
every coordination game on a directed graph in which all strongly
connected components are simple cycles is c-weakly acyclic.  Further,
in \cite{SW16} weighted open chains of cycles, closed chains of
cycles, and simple cycles with appropriate cross-edges were
considered.

For some of these classes of graphs some problems remain open, for
instance the existence of finite c-improvement paths for weighted open
chains of cycles. A rigorous presentation of the proofs of weak
acyclicity and c-weak acyclicity for the corresponding coordination
games is lengthy and quite involved.  We plan to present them in a
sequel paper. Finally, we believe that the following generalisation of
several of our results is true.

\begin{conjecture}
Coordination games on graphs with all nodes of indegree $\leq 2$ are c-weakly
acyclic. 
\end{conjecture}

Extensive computer simulations seem to support this conjecture.
However, our techniques do not seem to adapt easily to this bigger
class of graphs.

Next, by Nash's theorem, a mixed strategy Nash equilibrium always exists in coordination games irrespective of the underlying graph structure.
However, the complexity of finding one is an intriguing open problem.
This problem is known to be {\sc PPAD}-hard for various restricted classes of polymatrix games \cite{CD11,rubinstein2018inapproximability} 
(so it is unlikely to be solvable in polynomial time), 
but generalising this result to coordination games will be very challenging due
to the special structure of players' payoffs.
Still, we conjecture that this is indeed possible.

\begin{conjecture}
Finding a mixed Nash equilibrium in coordination games is a {\sc PPAD}-hard problem.
\end{conjecture}

Finally, note that in Section \ref{sec:directed-complexity} 
we assumed that all weights of the graph edges are natural numbers. It is
known that allowing weights to be rational may change the complexity
of the studied computational problem, e.g., the well-known knapsack
and partition problems become strongly NP-complete
\cite{wojtczak2018strong}. However, most computational problems for
coordination games with rational weights can be reduced in polynomial
time to the same problem for coordination games with integer weights
by simply multiplying all the weights by the least common multiple of all the
weights' denominators. This results in an exponential blow-up of value of the numbers, but only 
in a polynomial increase in their size
when they are represented in the standard binary notation. 

It is easy to see that a joint strategy is a Nash equilibrium or a strong Nash equilibrium
in the original game if and only if it is in the new game with the integer weights. So as
long as such a transformation results in a
coordination game of the type listed in Table \ref{fig:summary}, we get a polynomial time
algorithm for finding a Nash equilibrium or a strong Nash equilibrium in the original game. 
In particular, these problems for the coordination games with only two colours or on DAGs
can always be solved in polynomial time even when the weights are
rational. Notice that the problem of checking for the existence of a Nash equilibrium in a coordination games with rational weights is still in NP (simply guess a joint strategy
and check whether it is a Nash equilibrium) and at the same time it is NP-hard
as we already established it for the coordination games with the weights
equal to 0 or 1. So this problem is strongly NP-complete.

\section*{Appendix}
\NI
We provide here proofs of Lemmata \ref{lem:grades} and \ref{lem:mu-progress}.
\II

\lemgrades*

\textit{Remainder of the proof of Lemma \ref{lem:grades}.} To complete the proof of Lemma \ref{lem:grades} we provide a justification of the changes of the grades in Figures \ref{tab:first}, \ref{tab:umn}, \ref{tab:mn}, \ref{tab:qm}, and
\ref{tab:last}.

\NI
\begin{itemize}
    \item Figure \ref{tab:first}.

\NI
\textbf{Case 1}. The initial grade of $\calC_j$ is \mn.
This corresponds to the situation at the beginning of \emph{Phase 2}
in the proof of Theorem \ref{thm:TARK15a} when exactly one node has a
bonus.  This phase starts with the node $\nd{j}{2}$ and ends after at
most $n-1$ steps. So the colour of $\nd{j}{1}$ is not modified and
consequently the payoff to the down-link node $\nd{j+1}{k}$ of
$\calC_{j+1}$ is not modified.  Further the new grade of $\calC_j$ can
be either $\pl$ or $\upl$ depending whether at the end of this phase
the colours of $\nd{j}{v}$ and $\nd{j}{1}$ differ.  

\NI
\textbf{Case 2}. The initial grade of $\calC_j$ is \umn.
The reasoning is the same as in \textbf{Case 1}.  However, the colour
of $\nd{j}{v}$ is now not modified. The reason is that the only colour
that is propagated is that of $\nd{j}{1}$ and initially it is also the
colour of $\nd{j}{v}$. So the new grade of $\calC_j$ is now $\upl$.

\NI
\textbf{Case 3}. The initial grade of $\calC_j$ is \qm.
This corresponds to the situation at the beginning of \emph{Phase 1}
in the proof of Theorem \ref{thm:TARK15a} when exactly one node
has a bonus.  The constructed
improvement path ends after at most $2n-1$ steps, so in the process
the colour of $\nd{j}{1}$ can change.  If it does, then the grade of the
cycle $\calC_{j+1}$ can change arbitrarily.  In particular, it
can become \upl or \umn if the down-link node of $\calC_{j+1}$ is
$\nd{j+1}{v}$.
Further the new grade of $\calC_j$ can be either $\pl$ or $\upl$, for the same reasons
as in \textbf{Case 1}.

\NI
\item Figure \ref{tab:umn}.

The assumption that the grade of $\calC_j$ is initially \umn means
that initially the colours of $\nd{j}{1}$ and its predecessor
$\nd{j}{v}$ in this cycle are the same.  Then the construction in line
6 of the improvement path for the considered coordination game for
$\calC_j$ with bonuses for the link nodes corresponds to any update
scenario presented in {\em Phase 2} of the proof of Theorem
\ref{thm:cycle-2bonuses} that starts with \emph{i}.  There are six
such scenarios to consider.

\NI
\textbf{Case} \ci.
This means that the propagation of the colour of the up-link node of
$\calC_j$ stops before the down-link node of $\calC_j$ is reached.  So
the improvement path constructed in line 6 does not change the colours
of the link nodes of $\calC_j$ and of the predecessor $\nd{j}{v}$ of
the up-link node $\nd{j}{1}$.  Hence the grades of $\calC_{j-1}$ and
$\calC_{j+1}$ remain unchanged and the grade of $\calC_{j}$ becomes
\upl.

The remaining cases consider the situations in which the down-link
node of $\calC_{j}$ switches to another colour.  We now claim that in
these cases the grade of $\calC_{j-1}$ is initially \pl. Indeed, if
this grade is initially \upl, then the payoff to the up-link node
$\nd{j-1}{1}$ of $\calC_{j-1}$ is $\geq 1$.  But $\nd{j-1}{1}$ is
also the down-link node of $\calC_{j}$, so the claim follows by Lemma
\ref{lem:payoff}.

\NI
\textbf{Case} \cii.
This means that the propagation of the new colour of the up-link node of
$\calC_j$ stops between the down-link and up-link nodes of $\calC_j$
and that the down-link node adopted the colour of the up-link node. So
the improvement path constructed in line 6 does not change the colours
of $\nd{j}{1}$ and its predecessor $\nd{j}{v}$.

Hence the grade of $\calC_{j}$ becomes \upl and the grade of
$\calC_{j+1}$ remains unchanged.  On the other hand, the grade of
$\calC_{j-1}$ can remain unchanged or change from \pl to \mn, \upl of
\umn because of the new colour of the up-link node $\nd{j-1}{1}$ of
$\calC_{j-1}$.

\NI
\textbf{Case} \cio.
This means that the propagation of the colours stops between the
down-link and up-link nodes of $\calC_j$ but now the down-link node
(so $\nd{j-1}{1}$) adopted the colour of its predecessor $\nd{j-1}{v}$
in $\calC_{j-1}$.  So as in the previous case the grade of
$\calC_{j+1}$ remains unchanged.

However, the grade of $\calC_{j}$ can now also become \pl if this
propagation of the colours changes the colour of the predecessor
$\nd{j}{v}$ of the up-link node $\nd{j}{1}$.  Further, the grade of
$\calC_{j-1}$ now changes from \pl to \umn or \upl because the new
colour of $\nd{j-1}{1}$ is now the colour of $\nd{j-1}{v}$ and as a
result the node $\nd{j-1}{2}$ can now become the only node that does
not play a best response.

\NI
\textbf{Case} \cioi.
This means that the propagation of the colours now stops between the
up-link and down-link nodes of $\calC_j$ but now the down-link node
(so $\nd{j-1}{1}$) adopted the colour of its predecessor $\nd{j-1}{v}$
in $\calC_{j-1}$ and subsequently the up-link node $\nd{j}{1}$ of
$\calC_j$ adopted the colour of its predecessor $\nd{j}{v}$ in
$\calC_{j}$.  So the grade of $\calC_j$ now becomes \upl.

Further, the grade of $\calC_{j-1}$ now changes from \pl to \umn or
\upl for the same reasons as in the previous case.  Finally, the grade
of $\calC_{j+1}$ can now change arbitrarily for the same reasons as in
Case 3 concerning Figure \ref{tab:first}.

\NI
\textbf{Case} \cioo.
This case is similar to the previous one, with the difference that in
the second round of the propagation of the colours the up-link node
$\nd{j}{1}$ of $\calC_j$ adopted the colour of its predecessor in
$\calC_{j+1}$ instead of the colour of its predecessor $\nd{j}{v}$ in
$\calC_{j}$.  Consequently, the grade of $\calC_j$ now becomes \pl.
Further, the grade of $\calC_{j-1}$ can now change from \pl to \umn or
\upl, while the grade of $\calC_{j+1}$ can now change arbitrarily, both for
the same reason as in the previous case.

\NI
\textbf{Case} \ciooi.
This case cannot occur. Indeed, it would imply that the down-link node
in $\calC_j$ first switches to the colour of its predecessor in
$\calC_{j-1}$ and later switches to different colour. But the second
switch is not possible due to Lemma~\ref{lem:payoff}.



\NI
\item Figure \ref{tab:mn}.

The assumption that the grade of $\calC_j$ is initially \mn means that
initially the colours of $\nd{j}{1}$ and its predecessor $\nd{j}{v}$
in this cycle differ.  Then the construction in line 6 of the
improvement path for the considered coordination game for $\calC_j$
with bonuses for the link nodes corresponds to any update scenario
presented in {\em Phase 2} of the proof of Theorem
\ref{thm:cycle-2bonuses} that starts with \emph{o}.  There are four
such scenarios to consider.

\NI
\textbf{Case} \co.
The reasoning is the same as in \textbf{Case} \ci above with the
difference that the grade of $\calC_j$ becomes now \pl as the colours
of $\nd{j}{1}$ and $\nd{j}{v}$ do not change and hence remain different.

In the remaining cases the grade of $\calC_{j-1}$ is initially \pl for the
reasons given after \textbf{Case} \ci above.

\NI
\textbf{Case} \coi.
This case is analogous to \textbf{Case} \cii above. In particular, the
improvement path constructed in line 6 does not change the colours of
$\nd{j}{1}$ and its predecessor $\nd{j}{v}$.
Hence the grade of $\calC_{j}$ becomes \pl and the grade of
$\calC_{j+1}$ remains unchanged, while the grade of $\calC_{j-1}$ can
remain unchanged or change from \pl to \mn, \upl of \umn.

\NI
\textbf{Case} \coo.
This case is analogous to \textbf{Case} \cio above.  So, as in that
case, the grade of $\calC_{j+1}$ remains unchanged and the grade of
$\calC_{j-1}$ now changes from \pl to \umn or \upl.  However, the
grade of $\calC_{j}$ can now also become \upl if this propagation of
the colours changes the colour of $\nd{j}{v}$ to the colour of its
successor $\nd{j}{1}$.

\NI
\textbf{Case} \cooi.
This case is analogous to \textbf{Case} \cioi above.  So, as in that
case the grade of $\calC_j$ now becomes \upl, the grade of
$\calC_{j-1}$ changes from \pl to \umn or \upl, and the grade of
$\calC_{j+1}$ can change arbitrarily.

\NI
\item Figure \ref{tab:qm}.

This case corresponds to the situation at the beginning of \emph{Phase 1}
in the proof of Theorem \ref{thm:cycle-2bonuses}.  The constructed
improvement path ends after at most $3n$ steps, so in the process the
colour of $\nd{j}{1}$ can change.
Therefore, as in \textbf{Case 3} concerning Figure \ref{tab:first},
the grade of the cycle $\calC_{j+1}$ can change arbitrarily,  while
the grade of $\calC_j$ can become either $\pl$ or $\upl$.

Finally, if initially the grade of $\calC_{j-1}$ is \pl, then as in
\textbf{Case} \cii, its grade can remain unchanged or change to \mn,
\upl of \umn. Further, if initially this grade is \upl, then 
by the argument used in the proof of Lemma~\ref{lem:upl} the grade does not change.

\NI
\item Figure \ref{tab:last}.

We reduce the analysis for this case to the previous three cases by
extending the open chain with a new cycle $\calC_{m+1}$ in which all
new nodes have to their disposal colours that all differ from the
colours available to the nodes of $\calC_{m}$.  Then in Algorithm
\ref{alg:necklace} the bonus function for the up-link node of
$\calC_{m}$ is always 0 on the colours available to it, and
consequently for $j = m$ the improvement path constructed in line 6 of
Algorithm \ref{alg:necklace} is the same as for the original open
chain.  So for the case when $j=m$ we can use Figures \ref{tab:umn},
\ref{tab:mn}, and \ref{tab:qm} with the last columns always omitted.
This yields Figure \ref{tab:last}.

A perceptive reader can inquire why the row corresponding to the case
\cioo is missing.  The reason is that it deals with the situation when
the up-link node of $\calC_{j}$ switches to an outer colour, i.e, a
colour of its predecessor in $\calC_{j+1}$. But for $j=m$ this
cannot happen by the choice of the colours for the new nodes.
\qed

\II

We use below the following observation. 

\begin{claim}\label{x:observation-monotone}
  Let $s$ and $s'$ be two joint strategies such that
  $\mu(s) = (\guard(s), 0, |\prefix(s)|, -\nbr(s))$,
  $\guard(s) \leq \guard(s')$ and $|\prefix(s)| < |\prefix(s')|$. Then
  $\mu(s) <_{lex} \mu(s')$ holds.
\end{claim}

\proof{\sc Proof.}
Either $\mu(s') = (\guard(s'), 0, |\prefix(s')|, -\nbr(s'))$ or
$\mu(s') = (\guard(s'), 1, \ldots)$ and in both cases $\mu(s) <_{lex} \mu(s')$ holds. 
\qed
\end{itemize}
\endproof

\lemmuprogress*
\proof{\sc Proof.}
We check using Lemma \ref{lem:grades} that $\mu(s)$ increases
w.r.t.~the lexicographic ordering $<_{lex}$ each time one of the
updates presented in Figures \ref{tab:first}, \ref{tab:umn},
\ref{tab:mn}, \ref{tab:qm}, and \ref{tab:last} takes place.  So
throughout the analysis we assume that $j = \nbr(s)$. Let $s'$ denote
the new joint strategy computed in line 8 of the algorithm.
Lemma \ref{lem:upl} implies that $\guard(s) \leq \guard(s')$.  Further,
thanks to the definition of $\mu(s')$ we can assume that $s'$ is not a
Nash equilibrium. We consider each figure separately.


\begin{itemize}
\item Figure \ref{tab:first}.

\NI  
Then $j=1$ and $\guard(s) = 0$.

\NI
\textbf{Case 1}. The new grade of $\calC_{j}$ is \upl.
Then $\guard(s) < \guard(s')$ and hence $\mu(s) <_{lex} \mu(s')$.


\NI
\textbf{Case 2}. The new grade of $\calC_{j}$ is \pl.


\NI
\emph{Subcase 1}. $\mu(s) = (\guard(s), 1, 0,  -\nbr(s))$.

Then the initial grade of $\calC_{j}$ is \mn and $\prefix(s)$ contains
\upl, say at position $h$. Hence $\prefix(s')$ also contains \upl at
position $h$ and consequently $h \leq \guard(s')$.  But
$\guard(s) = 0$, so $\mu(s) <_{lex} \mu(s')$.

\NI
\emph{Subcase 2}. $\mu(s) = (\guard(s), 0, |\prefix(s)|, -\nbr(s))$.

If the initial grade of $\calC_{j}$ is \mn, then
$|\prefix(s)| < |\prefix(s')|$ since by assumption $s'$ is not a Nash
equilibrium. Otherwise the initial grade of $\calC_{j}$ is \qm and
then $|\prefix(s)| = 1$ by the definition of $\prefix(s)$, while
$1 < |\prefix(s')|$. So in both cases by Claim~\ref{x:observation-monotone}: 
$\mu(s) <_{lex} \mu(s')$.

\NI
\item Figure \ref{tab:umn}.

\NI  
By definition $\mu(s) = (\guard(s), 1, 0, -\nbr(s))$.

\NI
\textbf{Case 1}. The new grade of $\calC_{j-1}$ is \pl.
Then the case \ci or \cii applies and hence the new grade of
$\calC_{j}$ is \upl.  So $\guard(s) < \guard(s')$ and hence
$\mu(s) <_{lex} \mu(s')$.

\NI
\textbf{Case 2}. The new grade of $\calC_{j-1}$ is \upl.
Then $\guard(s) < \guard(s')$ and hence $\mu(s) <_{lex} \mu(s')$.

\NI
\textbf{Case 3}. The new grade of $\calC_{j-1}$ is \mn.
Then the case \cii{} applies and hence the new
grade of $\calC_{j}$ is \upl. So
$\mu(s') = (\guard(s'), 1, 0, -\nbr(s'))$.  But
$\guard(s) \leq \guard(s')$ and $-\nbr(s) < -\nbr(s')$, so
$\mu(s) <_{lex} \mu(s')$.  \II

\NI
\textbf{Case 4}. The new grade of $\calC_{j-1}$ is \umn.
Then $\mu(s') = (\guard(s'), 1, 0, -\nbr(s'))$ and $\mu(s) <_{lex} \mu(s')$
for the same reasons as in the previous case.

\NI
\item Figure \ref{tab:mn}.

\NI
\textbf{Case 1}. $\mu(s) = (\guard(s), 1, 0, -\nbr(s))$.
$\prefix(s)$ contains \mn at position $j$, so it contains \upl at some
position $h > j$. Moreover, by the definition of $\prefix(s)$ all
positions in it between $j$ and $h$ are \pl or \upl.

So if the new grade of $\calC_{j-1}$ is \pl or \upl, then
$j < \guard(s')$ and hence $\guard(s) < \guard(s')$ since
$\guard(s) < \nbr(s) = j$. So $\mu(s) <_{lex} \mu(s')$.
Otherwise the new grade of $\calC_{j-1}$ is \mn or \umn.  If it is
\mn, then $\prefix(s')$ contains \upl at the position $h > j-1$. So in
both cases $\mu(s') = (\guard(s'), 1, 0, -\nbr(s'))$.  But
$-\nbr(s) < -\nbr(s')$, so $\mu(s) <_{lex} \mu(s')$.

\NI
\textbf{Case 2}. $\mu(s) = (\guard(s), 0, |\prefix(s)|, -\nbr(s))$.
If the new grade of $\calC_{j-1}$ is \pl or \upl, then
$|\prefix(s)| < |\prefix(s')|$ since we assumed that $s'$ is not a
Nash equilibrium. So by Claim~\ref{x:observation-monotone}: $\mu(s) <_{lex} \mu(s')$.
If the new grade of $\calC_{j-1}$ is \mn or \umn, then $\guard(s) = \guard(s')$
and $|\prefix(s)| = |\prefix(s')|$ but
$-\nbr(s) < -\nbr(s')$, so $\mu(s) <_{lex} \mu(s')$.

\item Figure \ref{tab:qm}.

\NI  
By the definition $\prefix(s)$ ends with \qm, so $|\prefix(s)| = j$
and $\mu(s) = (\guard(s), 0, |\prefix(s)|, -\nbr(s))$.

If the new grade of $\calC_{j-1}$ is \pl or \upl, then
$j < |\prefix(s')|$, so by Claim~\ref{x:observation-monotone}: $\mu(s) <_{lex} \mu(s')$.
If the new grade of $\calC_{j-1}$ is \mn or \umn, then
$\guard(s) = \guard(s')$, $|\prefix(s)| \leq |\prefix(s')|$ and
$-\nbr(s) < -\nbr(s')$, so $\mu(s) <_{lex} \mu(s')$.

\item Figure \ref{tab:last}.

\NI  
The arguments for each case coincide with the arguments given for the
corresponding cases concerning Figures \ref{tab:umn},
\ref{tab:mn}, and \ref{tab:qm}.
\qed
\end{itemize}
%
\endproof

\subsection*{Acknowledgements}

  We are grateful to Mona Rahn and Guido Sch\"{a}fer for useful
  discussions and thank Piotr Sankowski and the referees of the
  preliminary, conference, versions for helpful comments.  First
  author was partially supported by the NCN grant
  2014/13/B/ST6/01807. The second author was supported by the
  Liverpool-India fellowship provided by the University of Liverpool
  and grant MTR/2018/001244. The last author was partially supported
  by the EPSRC grant EP/M027287/1.

\bibliographystyle{abbrv}


\end{document}